\documentclass[11pt,letter]{emulateapj}
\usepackage{graphicx}
\usepackage{epsfig}
\usepackage{amsmath}
\usepackage{amssymb}

\shorttitle{CGPS 408\,MHz Survey}
\shortauthors{Tung et. al.}

\begin{document}

\title{A High Resolution Survey of the Galactic Plane at 408\,MHz}

\author{A.K. Tung\altaffilmark{1,2}, R. Kothes\altaffilmark{1}, T.L.
Landecker\altaffilmark{1}, J. Geisb\"usch\altaffilmark{1,3}, D. Del
Rizzo\altaffilmark{1}, A.R. Taylor\altaffilmark{4,5}, C.M.
Brunt\altaffilmark{1,6},  A.D. Gray\altaffilmark{1}, S.M.
Dougherty\altaffilmark{1}}

\altaffiltext{1}{National Research Council Canada, Herzberg Programs in
Astronomy and Astrophysics, Dominion Radio Astrophysical Observatory, P.O. Box
248,  Penticton, British Columbia, V2A 6J9, Canada}         

\altaffiltext{2}{Department of Physics, University of Alberta, 4-181 CCIS,
Edmonton Alberta, T6G 2E1, Canada}

\altaffiltext{3}{Karlsruhe Institut f\"ur Technologie, PO Box 3640, 76021
Karlsruhe, Germany}

\altaffiltext{4}{Inter-University Institute for Data Intensive Astronomy, and
Department of Astronomy, University of Cape Town, Rondenbosch 7701, Republic of
South Africa}

\altaffiltext{5}{Department of Physics, University of the Western Cape, Republic
of South Africa}

\altaffiltext{6}{School of Physics, University of Exeter, Stocker Road, Exeter,
EX4 4QL, United Kingdom}

\begin{abstract} 
  {The interstellar medium is a complex `ecosystem' with gas constituents in the
  atomic, molecular and ionized states, dust, magnetic fields, and relativistic
  particles. The Canadian Galactic Plane Survey has imaged these constituents at
  multiple radio and  infrared frequencies with angular resolution of the order
  of arcminutes.}  {This paper presents radio continuum data at 408\,MHz over
  the area ${52^{\circ}}\leq{\ell}\leq{193^{\circ}}$,
  ${-6.5^{\circ}}\leq{b}\leq{8.5^{\circ}}$, with an extension to
  ${b}={21^{\circ}}$ in the range ${97^{\circ}}\leq{\ell}\leq{120^{\circ}}$,
  with angular resolution $2.8' \times 2.8'$\,cosec\thinspace$\delta$.} 
  {Observations were made with the Synthesis Telescope at the Dominion Radio
  Astrophysical Observatory as part of the Canadian Galactic Plane Survey. The
  calibration of the survey using existing radio source catalogs is described.
  The accuracy of 408-MHz flux densities from the data is 6\%.} {Information on
  large structures has been incorporated into the data using the single-antenna
  survey of Haslam (1982).} {The paper presents the data, describes how it can
  be accessed electronically, and gives examples of applications of the data to
  ISM research.}
\end{abstract}


\section{INTRODUCTION}
\label{intro}

The Canadian Galactic Plane Survey (CGPS) is a survey of the major constituents
of the interstellar medium (ISM), designed to capture the atomic gas (\ion{H}{1}
observed at a wavelength of 21.1\,cm), relativistic and ionized components
(radio continuum observed at 21.1 and 73.4\,cm), molecular gas (observed at a
wavelength of 2.6 millimetres), and dust (observed between 12.5 and 100
microns). The scientific rationale of the CGPS and many technical details of the
survey, including an outline of the survey procedure, can be found in
\citet{tayl03}. In this paper we present that part of the survey observed at
408\,MHz (wavelength 73.4\,cm) with the Synthesis Telescope at the Dominion
Radio Astrophysical Observatory (which we refer to as the DRAO ST, described by
\citealp{land00}). Radiation from the Milky Way at 408\,MHz is predominantly
synchrotron emission, but the effects of the ionized gas are also seen, so the
images portray emission from the relativistic and warm ionized components of the
ISM.

When a low-frequency channel was added to the DRAO ST in the early 1980s, the
408\,MHz frequency band was chosen because of the availability of the
\citet{hasl82} survey of the entire sky at that frequency. The Haslam data could
provide information on the largest structures, those to which the Synthesis
Telescope is less sensitive. This continued the DRAO practice of combining
aperture-synthesis data with single-antenna data (pioneered by \citealp{higg77})
in order to represent fully the spatial structures on the sky. As a result, the
images presented in this paper portray the Galactic emission on all scales from
the largest to the ${\sim}2.8'$ resolution limit of the telescope.

At 408\,MHz the telescope receives right-hand circular polarization only. It
correctly measures the total intensity of synchrotron emission, assuming a
negligible content of circular polarization, but is not sensitive to any
linearly polarized components.

The 408\,MHz system of the DRAO ST (\citealp{lo84,veid85}) has an
analog-to-digital converter (ADC) with 2-bit quantization, but only three of the
four possible levels are used, which yields a limited dynamic range. The sky
signal at 408\,MHz, especially from the plane of the Galaxy, is strong enough to
dominate the receiver noise; under these circumstances the ADC could be
operating outside its optimum range, leading to loss of sensitivity. The
telescope is therefore equipped with an automatic level-control (ALC) system to
keep the correlator input within the optimum range.  A side-effect of the ALC
system is that it renders ineffective the standard amplitude calibration of the
telescope, because the system gain is almost always different when the telescope
is observing the calibration source than when observing the survey field. As a
consequence, the calibration has to be restored in the data processing pipeline.
Since the ALC responds to the total power coming into the front end of the
receiver, this limitation can be overcome to some extent by evaluating all-sky
data such as the 408\,MHz all-sky survey of \citet{hasl82}. However, a more
precise re-calibration is desirable, preferably one based on one or more
well-recognized source catalogs, and that is a major topic in this paper. It is
important to note that the ALC system has no effect on the phases derived from
calibration observations, so image quality is not impaired.


\section{OBSERVATIONS AND DATA PROCESSING}
\label{obs}

Table~\ref{table:specs} presents the parameters of the survey. In this section we
describe those aspects of the observations and data processing that are
particularly germane to the 408\,MHz survey data. 

\begin{table}
\caption[]{Survey properties of the CGPS relevant to this work.}
\begin{center}
\begin{tabular}{ll}
\hline
Coverage                     & ${52^{\circ}} < {\ell} < {193^{\circ}},
                               {-6.5^{\circ}} < {b} < {8.5^{\circ}}$ \\
                             & ${97^{\circ}} < {\ell} < {120^{\circ}},
                               {5.0^{\circ}} < {b} < {21^{\circ}}$ \\
Total survey area            & 2204 square degrees \\
Number of fields             & 448 \\
Spacing of field centers     & $117'$ hexagonal grid \\
Antenna primary beam         & $332.1'$ FWHM \\
Dates of observations        & 1995.3 to 2009.2 \\
Center frequency             & 408\,MHz \\
Bandwidth                    & 3.5\,MHz \\
Polarization products        & Right-hand circular polarization \\
Angular resolution           & $2.8' \times 2.8'$\,cosec\thinspace$\delta$ \\
Sensitivity                  & 3 mJy/beam rms \\
Typical noise in mosaicked   & \\
images                       &  
${0.76}\thinspace{\rm{sin{\thinspace}\delta}}$~K \\
Source of single-antenna data & \citet{hasl82} \\
\hline
\end{tabular}
\end{center}
\label{table:specs}
\end{table}

The DRAO ST employs relatively small antennas (diameter $\sim 9$\,m). The
field-of-view is wide, and the small antennas permit the telescope to sample
interferometer baselines as short as 12.9\,m, corresponding to spatial
structures as large as ${\sim}3^{\circ}$ at 408\,MHz. Information on larger
structures is provided by data from single-antenna telescopes incorporated into
the imaging process. Sampling of the ($u,v$) plane is also very thorough,
covering from 12.9\,m to 604.3\,m in steps of 4.3\,m (plus one sample at
617.1\,m). The telescope therefore has excellent sensitivity to low-level
extended emission, unlike many other aperture synthesis telescopes. Furthermore,
the dense and regular sampling of the ($u,v$) plane moves the first grating lobe
out to a radius of $9.8^{\circ}$, beyond the primary beam area. The thermal
noise on an individual field is 5\,mJy/beam at 408\,MHz, but this level is
usually not attained because the images are confusion limited (we discuss
confusion in this survey in Section~\ref{conf}).

The survey was carried out as a series of pointings of the telescope, with data
from each pointing processed into a separate image.  The pointings were placed
on a hexagonal grid, with spacing between field centers of $117'$, chosen to
give good sampling at 1420\,MHz where the antenna beamwidth is $107.2'$ FWHM.
The individual fields were therefore very closely spaced relative to the
408\,MHz beam of $332.1'$ FWHM. The data processing pipeline used software from
the DRAO Export Package (\citealp{higg97,will99}). At the end of the data
processing pipeline the 448 images were mosaicked together.  The data are
released as $15^\circ \times 15^\circ$ individual mosaics (see
Section~\ref{pres}). 

Before mosaicking, single-antenna data from \citet{hasl82} were incorporated
into the image of each field. The single-antenna data were transformed to the
visibility (${u,v}$) plane. Visibilities were divided by a Gaussian function of
FWHM equivalent to a baseline of 20~m, the transform of the profile of the
51-arcminute beam of the Haslam data. The DRAO ST images were similarly
transformed to the (${u,v}$) plane. The two datasets were merged in the
(${u,v}$) plane using normalized functions to taper each in the overlap zone of
12.9 to 30.0~m, with value 0.5 at the overlap radius of 21.4~m. The amplitude
scales of single-antenna and aperture-synthesis data are matched within 10\%.
Low-level striping of amplitude a few K is sometimes evident in the data. This
is introduced into the images from the \citet{hasl82} data, consistent with the
zero-level uncertainty of ${\pm}3\thinspace{\rm{K}}$ quoted for that survey.

\section{CALIBRATION}
\label{cal}

The DRAO ST observations were initially calibrated using short observations of
3C\,147 and 3C\,295, and, less frequently, 3C\,48, assuming flux densities at
408\,MHz of these unpolarized sources{\footnote{\citet{perl13} provide
information on the polarization properties of these sources at low
frequencies.}} of 48.0, 54.0, and 38.9\,Jy respectively
\citep{tayl03}. We refer to images calibrated in this way as {\textit{raw
data}}. The operation of the ALC system made the amplitude calibration of the
survey unreliable, so a process of ``registration'' was developed to tie the
amplitude scale of the survey to other surveys. Reference surveys were sought
with comparable angular resolution at frequencies above and below 408\,MHz and 
as close as possible to it. Initially the NRAO VLA Sky Survey (NVSS) at 1.4\,GHz
\citep[NVSS,][]{cond98} and the Cambridge 7C(G) survey at 151\,MHz
\citep{vess98} were chosen,  but uncertainties of order 15\% remained in the
observations processed in this way {\citep{tayl03}}. Uncertainty arose from
undetected spectral curvature of the sources chosen as calibrators, and the fact
that the NVSS was tied to the flux scale of \citet{baar77} while 7C(G) was tied
to the flux scale of \citet{roge73}. In addition, 7C(G) does not cover the
entire CGPS area.

With more surveys now available at low frequencies, we have re-calibrated the
CGPS 408-MHz survey using the NVSS, the 74\,MHz VLA Low-frequency Sky Survey
\citep[VLSS,][]{cohe07} and the 365\,MHz Texas Survey of Radio Sources
\citep[Texas,][]{doug96}. These three surveys are tied to the Baars scale or can
be converted to it, and the survey coverage includes almost the entire CGPS
observation area. Using three reference surveys enabled us to eliminate sources
with complex spectra from the calibration process. The WENSS survey at
325\,MHz \citep{reng97} would be useful in this work, but it does not cover the
full CGPS area.

\subsection{Calibration Source Selection}
\label{sel}

Source selection followed a multi-step procedure to select only bright, compact
sources with simple power-law spectra that appeared in all three catalogs.
First, we rejected sources which were flagged in the various catalogs as
extended, complex in structure, or variable, or which had high noise residuals.
Sources with signal-to-noise ratio less than 5 in any one of the three catalogs
were also rejected. Next, it was necessary to determine that all three catalogs
were referring to the same source. Using the NVSS catalog co-ordinates as a
reference, for their lower uncertainty, we selected only sources with
counterparts within $3\sigma$ position errors in both the Texas Survey ($<5''$)
and VLSS ($<60''$).  Of these, we kept only sources that had monotonically
decreasing flux density with increasing frequency, consistent with a non-thermal
power-law spectrum.  This left 13471 sources in the region where all three
surveys overlap ($-30^\circ < \delta < 71.5^\circ$), of which 891 were
within the area covered by the CGPS.

A deeper analysis of the source spectra was necessary to eliminate those with
complex or curved spectra that could not be adequately modelled by the power-law
relation 
\begin{equation} S = S_0 \left(\frac{\nu}{\nu_0}\right)^{\alpha}
\end{equation} 
where $S_0$ is the flux density at $\nu_0 = $1000\,MHz and $\alpha$ is the
spectral index.  Power-law spectra were fitted to every one of the 13471
sources, minimizing a weighted $\chi^2$.  The best-fit spectra were used to
calculate the flux densities of each potential calibration source at 408\,MHz. 
The fitting parameters, including the errors of individual variables and their
covariances, were derived to calculate the total error in the flux density. In
the case of power-law spectra, the total error was calculated from %
\begin{equation} dS^2 = dS_0^2 \left(\frac{dS_0}{dS}\right)^2 +  d\alpha^2
\left(\frac{d\alpha}{dS}\right)^2 +  cov(\alpha,S_0)
\left(\frac{dS_0}{dS}\right) \left(\frac{d\alpha}{dS}\right). \end{equation} %
In the following discussion we refer to the 408\,MHz values obtained in this way
as {\textit{interpolated flux densities}}.

We found that the best-fit power-law spectrum tended to systemically
over-estimate the 74\,MHz (VLSS) flux densities. The causes could include
scale errors in the basic surveys, synchrotron aging affecting the high-energy
end of the electron spectrum, or free-free absorption in ionized gas near the
Galactic plane affecting measured flux densities at the lowest frequencies. To
study spectral complexity around 408\,MHz we drew additional information
from the final non-redundant catalog from the Cambridge 7C survey at 151\,MHz
\citep{hale07} and the 325\,MHz Westerbork Northern Sky Survey
\citep[WENSS,][]{reng97}. The 7C survey is tied to the flux-density scale of
\citet{roge73}; since there is no direct conversion factor that transfers the
scale of \citet{roge73} to that of \citet{baar77}, this analysis serves only as
a method to assess the spectral complexities of the catalog sources without
determining reliable source spectra. In the limited area where these two surveys
overlap with the initial three we used, we found 2575 sources that met our
criterion of $3\sigma$ position coincidence. We fitted straight lines and
polynomials of order two and three in the log-frequency log-flux-density domain,
using all five frequencies for those sources. 

To determine the preferred model for each spectrum we calculated the Bayesian
Information Criterion (BIC) as  
\begin{equation}   BIC = {{\chi}_s}^2 + {k}\log{n},   
\end{equation}  
where the values of ${{\chi}_s}^2$ could easily be computed from the best fit
parameters, $k$ is the number of parameters we are fitting to, and $n$ is the
number of data points, 5 in this analysis. The preference for a model was
considered \textit{positive} if it yielded a BIC value a factor of 2 or more
lower than was produced by other models, and \textit{strong} if that factor was
6 or more lower \citep{kass95}.  Of the 2575 sources, a total of 972
showed a positive preference for the second-order polynomial model, of which 294
showed a strong preference.

When we compared results for these sources with the fits made using only the
NVSS, Texas Survey, and VLSS, we found that ${{\chi}_s}^2$ was a sufficient
discriminator: of sources with ${{{\chi}_s}^2} < 5$, only 10\% strongly
preferred a non-power-law spectrum. We thus consider that sources in the
three-catalog fit with ${{{\chi}_s}^2} < 5$ to be \textit{good} calibrators, and
sources with ${{{\chi}_s}^2} < 3.5$ to be \textit{excellent} calibrators.

Of the 891 potential calibrators in the CGPS area, 417 are good calibrators and
357 of those are excellent. On average, we found 5 to 10 good or excellent
calibration sources present in each 408\,MHz CGPS field, with the exception of
the Cas-A and Cyg-A regions, where the densities of sources in all three
catalogs drop. We have compiled a list of good and excellent (as defined above)
calibration sources that covers almost the entire sky accessible to the DRAO ST,
and that list is available as a general utility for the telescope.  We refer to
this list as the \textit{calibration database}{\footnote{The calibration
database will be available with the on-line version of this paper, and will also
be available at the Canadian Astronomy Data Centre
(www.cadc-ccda.hia-iha.nrc-cnrc.gc.ca/en/cgps) when the paper is published}}.
Table~\ref{table:cal_src} provides an example for the first three sources in
this database. For each source the table gives the source designation and
position from the NVSS catalog, the flux densities with errors as observed by
the three catalogs, and the interpolated 408\,MHz flux density, with error.  The
table also lists the fitting parameters, including flux at the reference
frequency (1000\,MHz), spectral index, covariance, and weighted $\chi^2$. 
Table~\ref{table:cal_src} has 7686 entries, all of them meeting our criterion of
good calibrators, and of these 6614 are excellent. We do not give information on
spectral curvature in Table~\ref{table:cal_src} because it is not needed to
achieve an accurate calibration. For each 408-MHz flux density derived from a
simple power-law fit we give the spectral index and the probable error in flux
density.

\subsection{Calibrating Single Fields}

Calibration sources were selected from the calibration database for each of the
448 individual survey fields (see Section~\ref{obs}). Sources were chosen for a
given field if they lay within $4.1^{\circ}$ of the field center, where the
primary beam is above 20\% of its on-axis level. The flux densities of the
corresponding sources were then extracted from the raw CGPS image, prior to
addition of single-antenna data. All source fitting used the routine
{\tt{fluxfit}} from the DRAO export package. Each compact source was fitted with
a two-dimensional Gaussian function above a twisted-plane background; the
background region was always three times the synthsized beam in each of
declination and right ascension. The calibration factors of individual fields
were derived by fitting a straight line, anchored at the origin, on a scatter
plot between raw flux densities and the interpolated flux densities of the
calibration sources within the field. The model that we fitted to the scatter
plot is described by
\begin{equation} 
S_{\rm i} = a S_{\rm raw},  
\end{equation} 
where $S_{\rm i}$ is the interpolated flux density, $S_{\rm raw}$ is the
raw flux density, and $a$ is the calibration factor. Since there are
errors in both the interpolated and raw flux densities, we extended the
least-squares method to two dimensions, where the algorithm minimized weighted
residuals in both x and y directions. For a data point at $(x,y)$ with errors
$(\Delta x_i, \Delta y_i)$, there is a corresponding point on the trend line
where a perpendicular line that goes through the data point intersects the
trend line at $(x_0,y_0)$. The intersection point satisfies
\begin{equation} 
ax_0 = y_0 = -\frac{x_0}{a} + y + \frac{x}{a}.
\end{equation} 
After reduction, this equation yields
\begin{equation} 
x_0 = \frac{ay + x}{a^2 + 1} 
\end{equation} 
and 
\begin{equation} y_0 = \frac{a^2y + ax}{a^2 + 1}. 
\end{equation} 

The perpendicular distance, $d$, from the data point $x, y$ to the point $x_{0}, y_{0}$ is
\begin{equation}
d = \sqrt{(x - x_{0})^{2} + (y - y_{0})^{2}} = \sqrt{\frac{(y - ax)^{2}}{a^{2} 
+ 1}}.
\end{equation}
Assuming independent errors on $x$ and $y$, propagation of errors onto $d$ 
yields
\begin{equation}
{\Delta}d = \sqrt{\frac{a^{2}({\Delta}x)^{2} + ({\Delta}y)^{2}}{a^{2} + 1}} .
\end{equation}
The contribution to the ${\chi}^{2}$ of the overall fit from each data point is
the square of the perpendicular distance in units of its error. When summed over
all  $i = 1 ... n$ data points, we have
\begin{equation}
{{\chi}_{fit}}^{2} = 
\sum_{i=1}^n \frac{d_{i}^{2}}{({\Delta}{d_i})^{2}} = \sum_{i=1}^n
\frac{(y_{i} - ax_{i})^{2}}{a^{2}({\Delta}{x_i})^{2} + ({\Delta}{y_i})^{2}} .
\end{equation}

Note that this can be viewed as the result of {\it active} scaling of the
$x$ data, and associated errors, by the calibration factor $a$. Minimizing
${{\chi}_{fit}}^{2}$ therefore relies on the statistical equivalence of $y$ and
$ax$, with weighting provided by their combined uncertainty in the denominator.
Since the ${{\chi}_{fit}}^{2}$ expression is non-linear in $a$, we minimized it
by a grid search to deduce the best-fitting calibration factors.

\begin{figure} 
\begin{center} 
\includegraphics[bb = 65 110 560 590,scale=0.45,clip=]{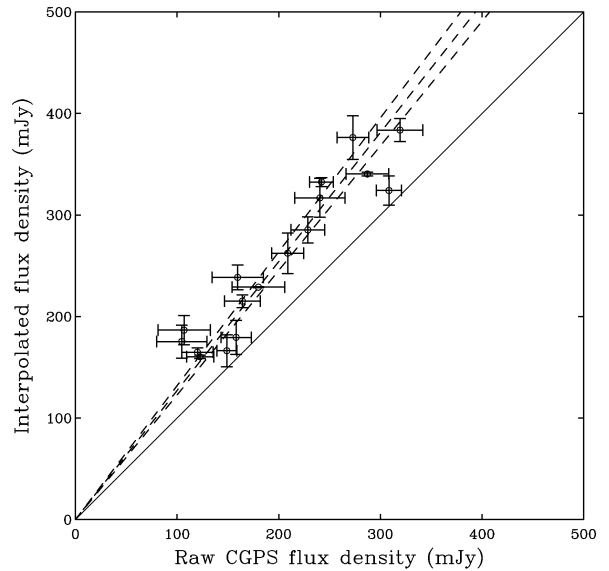} 
\caption{Illustrating the first pass of the calibration procedure, as applied to
field S6, centered at ${(\ell,b)}={(154.3^{\circ},2.6^{\circ})}$. Flux densities
for calibration sources, selected as described in the text, are plotted against 
their flux densities in the raw CGPS image. $1\sigma$ errors are shown. The
solid line  represents $y=x$ and the dotted lines show the best fit,
$y=1.2706x$, and fits  $1\sigma$ above, $y=1.3188x$, and below the best fit,
$y=1.2271$. The  calibration factor for this field is $1.2706\pm{3.5\%}$.} 
\label{S6} 
\end{center} 
\end{figure}

\begin{figure} 
\begin{center} 
\includegraphics[bb = 70 110 555 592,scale=0.45,clip=]{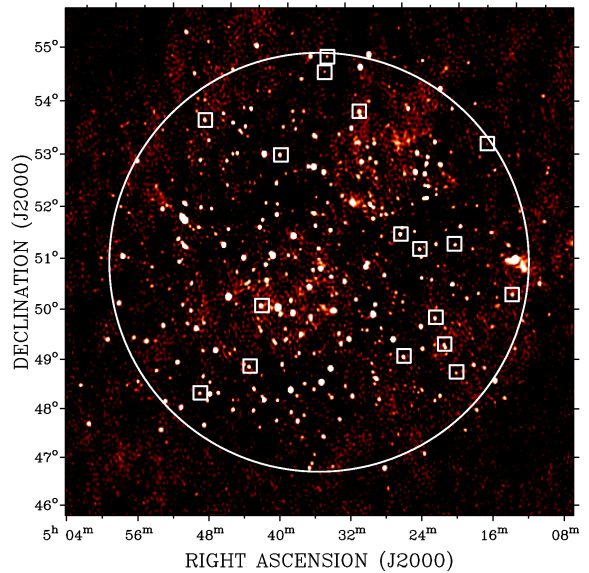} 
\caption{Raw data for the field S6, centered at
${(\ell,b)}={(154.3^{\circ},2.6^{\circ})}$. No correction has been made for the
primary beam of the antennas, and single-antenna data have not been incoporated
into this image. The color scale has been chosen to emphasize point sources in
the field. The white boxes enclose the 17 excellent calibration sources
in this field, all of which lie inside the 20\% level of the primary beam, shown
by the white circle. Calibration using these 17 sources is illustrated in
Figure~\ref{S6}.}
\label{S6-raw} 
\end{center} 
\end{figure}

The calibration fitting for a typical field is illustrated in Figure~\ref{S6}.
The first pass of the calibration procedure was conducted on fields with three
or more good calibration sources within the primary beam, out of which more than
half had three or more excellent calibration sources. The distribution of these
sources on the raw data for this field is shown in Figure~\ref{S6-raw}.

\begin{figure}
\begin{center}
\includegraphics[bb = 60 115 550 585,scale=0.45,clip=]{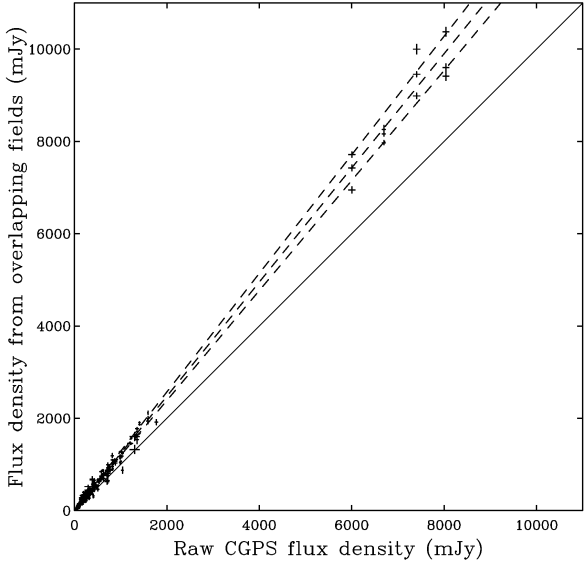}
\caption{Illustrating the second pass of the calibration procedure, applied
where a field contains an inadequate number of good calibrators, in this case
field C7, centered at ${(\ell,b)}={(119.6^{\circ},-2.2^{\circ})}$. Sources in
fields that overlap C7, calibrated in the first pass, are used as calibrators.
The flux densities of 152 such calibrators are plotted against the flux density
of the same sources in the raw images. Three separate flux densities are shown
for each source, measured in three separate overlapping fields. All source flux
densities are shown with $1\sigma$ error bars. The solid line represents $y=x$
and the dotted lines show the best fit, $y=1.2375x$, and fits  $1\sigma$ above,
$y=1.2848x$, and below the best  fit, $y=1.1919x$. The calibration factor for
this field is $1.2375\pm{3.8\%}$.}
\label{C7}
\end{center}
\end{figure}

\begin{figure}
\begin{center}
\includegraphics[bb = 100 115 495 505,scale=0.45,clip=]{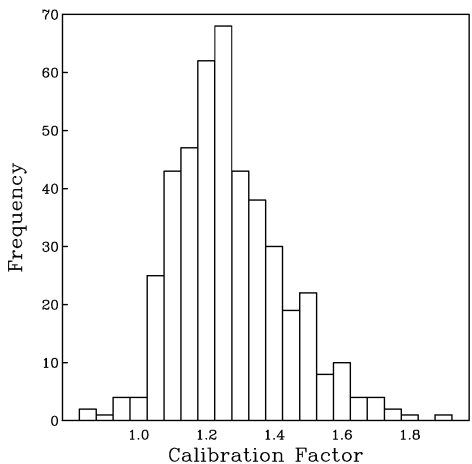}
\caption{Histogram of calibration factors derived for 438 out of the 445 fields
fields in the survey. Six fields near the strong sources Cas~A and Cyg~A are 
omitted. One field, in the vicinity of Cyg~A, has a calibration factor $>3.0$ 
(not shown).}
\label{histo}
\end{center}
\end{figure}

The second pass of calibration was conducted on fields that were considered to
contain an insufficient number of good calibration sources (less than three).
The calibration can be transferred to such fields from neighboring fields
because of the substantial overlap, 70\% by area, between adjacent
408\,MHz fields (see Section~\ref{obs}).

While some fields may have less than three \textit{excellent} calibration
sources, tens to hundreds of unresolved sources can be extracted from most
fields. Using the DRAO Export Package, we located and extracted bright point
sources, at least $3\sigma$ above background, from the uncalibrated field and
from surrounding calibrated fields; these sources were all above 50 mJy, and
were used to transfer the calibration from a calibrated field to an
overlapping one. All fields in this process were corrected for attenuation by
the primary beam.  The calibration was extended from the calibrated field to
an overlapping uncalibrated one by fitting a straight line to a plot where the
raw flux density is the abscissa and the flux density for the same source
obtained from an overlapping field is the ordinate. The process is illustrated
in Figure~\ref{C7}. If there was more than one overlapping field, all fields
were used in this process; in the example of Figure~\ref{C7} there were three
overlapping calibrated fields. The fitting followed the algorithm described
above for the first step of the calibration.

We established a hierarchy for calibrating fields that required this method.
Those uncalibrated fields with the most overlapping calibrated fields were
calibrated first, and then used as calibrated fields in support of other
uncalibrated fields.  The field that was selected and calibrated first would
preferably have all of its six surrounding fields calibrated. Where no such
field could be found, we selected fields surrounded by five calibrated fields to
calibrate next, and so on. If two or more fields had the same number of
surrounding calibrated fields, we calibrated them independently and returned to
the beginning of the selection procedure. This hierarchy transferred calibration
from the outer fields surrounding the troublesome regions towards the center and
ensured minimal propagation of calibration errors. Figure~\ref{C7} shows the fit
for a field that is surrounded by three calibrated fields.

Out of the 445 observed CGPS fields, 367 were calibrated directly using the
calibration sources, 72 were calibrated by matching the point sources between
calibrated and uncalibrated fields, and only six fields, with pointing centers
very close to Cas~A or Cyg~A, could not be calibrated by either of the methods
described. However, the phase centers of the fields are very closely spaced
compared to the field of view, and data for the areas immediately around Cas~A
and Cyg~A were simply taken from fields with slightly more distant phase
centers. Fields with phase centers very close to these strong sources were not
used in computing the final mosaics (it should be noted that areas around these
strong sources are artefact dominated and are generally not useful for most
analysis). Figure~\ref{histo} shows the distribution of calibration factors for
438 of the fields in the survey. Almost all calibration factors are larger than
1.0 because the strong extended emission along the Galactic plane has reduced
the telescope gain relative to the gain at higher latitudes where 3C\,147,
3C\,295, and 3C\,48 lie.

\section{ERROR ANALYSIS}

\subsection{Errors in Flux Calibration}

Two factors contribute to the calibration error of the 408-MHz survey.  First,
the calibration sources may have spectral curvature despite our efforts to
eliminate such sources. Second, there may be inconsistencies in the flux density
of sources within one field, arising either from instrumental errors or errors
in calibration of the telescope at the time of the observation of that field. We
consider these sources of error in turn.

Even though we reduced the number of sources that strongly prefer to be modelled
with complex spectra, we cannot completely avoid sources which have slight
curvature in their spectra. From the list of 7686 excellent and good calibration
sources, we compared the flux densities derived from best-fit spectra with those
recorded in the catalogs used at the three respective frequencies. The flux
densities agree very well with the NVSS catalog at 1400\,MHz, with an average
difference in flux of $\sim0.4\%$ between the interpolated and the catalog flux
densities. However, at 365\,MHz, the derived flux densities are on average
$\sim4\%$ lower than recorded by the Texas catalog, and at 74\,MHz the
interpolated flux densities are on average $\sim5\%$ higher than recorded by the
VLSS catalog. Interpolating to 408\,MHz, the systematic error due to spectral
curvature is $<4\%$. These results may indicate overall scale errors in one
or more of the three surveys that are the basis for the calibration, or point to
any of the other causes discussed in Section~\ref{sel}.

Second, we considered the internal field-to-field consistency of calibration. We
chose eight sources with flux densities between 1 and 2~Jy from various parts of
the survey. The fitting process that we used to calibrate the survey has its
best performance in this range of flux density. We measured the flux densities
of the eight sources from survey images after the scaling factor had been
applied and compared them with flux densities derived by interpolation among the
three catalogs, NVSS, Texas, and VLSS. The flux densities from the final survey
images were within 1 to 4.5\% of their interpolated flux densities, with an
average value $2.38\pm0.33\%$. Note that these sources were {\textit{not}}
calibration sources. On the basis of our examination of these two sources of
error we estimate the systematic errors in flux density for the survey to be
less than $6\%$.

\citet{perl16} have established a new flux-density scale between 50\,MHz and
50\,GHz based on VLA measurements. They give polynomial fits that define the
flux density of a number of sources as a function of frequency. It is difficult
to compare our calibration with this work because none of the \citet{perl16}
sources are within the CGPS area. The flux densities predicted for 3C\,147,
3C\,295 and 3C\,48 at 408\,MHz are within 1.3\% of the values used for initial
calibration of our data, quoted in Section~\ref{cal}, but that is not really
relevant to the problem. The best that we can do is to compare the
\citet{perl16} scale with the \citet{baar77} scale on which our calibration is
based. \citet{perl16} give ratios of their flux densities to values from
\citet{baar77} at low frequencies; averaged over 10 sources, the \citet{perl16}
scale is 0.5\% lower than the \citet{baar77} scale at 328\,MHz and 1.3\% higher
at 1488\,MHz. The agreement is well within the error that we have estimated for
our flux densities.

\subsection{Accuracy of Representation of Extended Structure}
\label{extended}

The flux scale of the survey is calibrated using point sources, as described
above, and we have a good understanding of the probable error in that scale. But
how accurately does the survey portray extended structure, and what is the
relative error between large and small scales?

The accuracy of the representation of extended structure depends partly on the
calibration accuracy of the Haslam data. The brightness temperature scale of the
Haslam survey is tied to the survey by \citet{paul62} which was calibrated by
reference to absolute standards of noise (resistors at known temperatures) and
an evaluation of antenna  gain from a measurement of the complete radiation
pattern of the telescope. This procedure, apparently done carefully, has
provided a good calibration for extended emission, relevant here, but not
necessarily for compact sources (for a discussion see \citealp{rema15}).

The process of combining the Haslam data with those from the DRAO ST is a fairly
conventional one: the two datasets are spatially filtered and added. The two
datasets blend smoothly over the central regions of the $(u,v)$ plane; details
are given in Section~\ref{obs}. The match between the two independent
calibrations has been checked in the range where the two datasets share $(u,v)$
plane coverage (which is large, covering at least 13 to 50 metres) and we
estimate that the amplitudes are matched within 10\%. Changing the balance
between the two datasets within this range makes a barely discernible difference
to the images shown in the following sections. 

\section{SOURCE CONFUSION}
\label{conf}

\begin{figure}
\begin{center}
\includegraphics[bb = 75 110 565 450,scale=0.45,clip=]{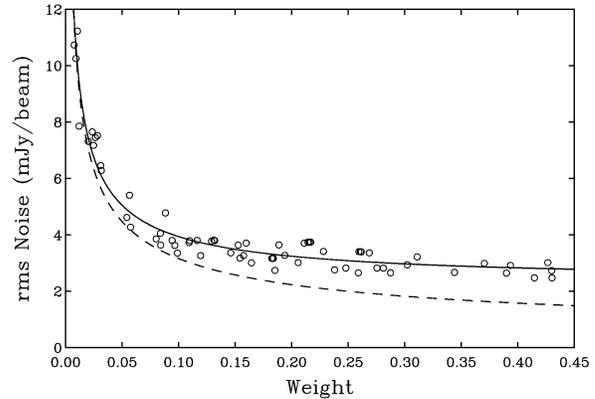}
\caption{The effects of source confusion on noise in a mosaic. See text for
description of the method used to generate this plot. The solid curve shows
noise in the presence of confusion while the dashed curve shows the noise
theoretically achievable without confusion.}
\label{noise}
\end{center}
\end{figure}

Source confusion, the additive effect of unresolved compact sources in the beam,
limits the sensitivity of this survey. We studied confusion effects using mosaic
C1, centred at ${(\ell,b)}=({142.3^{\circ},1.0^{\circ}})$; this mosaic is far
from any strong source, contains relatively little extended emission, and is
free from solar contamination or other interference. In this experiment we used 
only fields whose noise (after calibration) was less than 1\,K. We constructed
the mosaic 18 times, using a different number of fields in each trial, and
examined the decrease in noise level as more fields were incorporated. The
results are shown in Figure~\ref{noise}.

The correction for the primary beam function increases the noise across an
individual field away from the field center. The mosaicking algorithm sums the
contributions from individual fields, applying a weight, $g$, to each point that
depends on the noise level: ${g}$ is given the value ${1/{{\sigma}^2}}$ where
${\sigma}$ is the rms noise in the vicinity of that point.  Weight increases as
successive fields are incorporated into the mosaic, and as the mosaic builds up
the algorithm generates a weight map. Weight varies across the mosaic, depending
on the number and quality of the individual fields contributing in a particular
part of the mosaic, and weight is lower at the edges of the mosaic because of
attenuation by the primary beam of the telescope. To generate Figure~\ref{noise}
we identified small source-free patches at various places in the mosaic at
intermediate stages of the mosaic assembly, and therefore with different
weights. We plotted rms noise in the patch (in mJy/beam) against the value of
weight at the centre of that patch. In the absence of confusion the rms noise
should decrease as ${g}^{0.5}$. In the presence of confusion the noise will vary
as 
\begin{equation}   
{{\sigma}_{\rm{total}}^2} = {{{\sigma}_{\rm{confusion}}^2}+{1/g}}. 
\end{equation}   
In Figure~\ref{noise} we show a curve fitted to this equation as well as the
theroretical curve in the absence of confusion.

Our conclusion from the data in Figure~\ref{noise} is that the noise in an
individual field is about 5\,mJy/beam without confusion, rising to 5.5\,mJy/beam
with confusion. We achieve a sensitivity of about 3\,mJy/beam in the mosaics,
and in the absence of confusion would go about two times deeper. Our
estimate of the confusion limit is
${{\sigma}_{\rm{confusion}}}={2.4}${\thinspace}mJy/beam. We can compare this
with the  calculated confusion limit following the work of \citet{cond02}; from
the equations in that paper we obtain 3.5 mJy/beam for this particular field of
the survey. We regard this as satisfactory agreement.

\begin{figure*}
\begin{center}
\includegraphics[bb = 90 35 490 765,scale=0.85,angle=0,clip=]{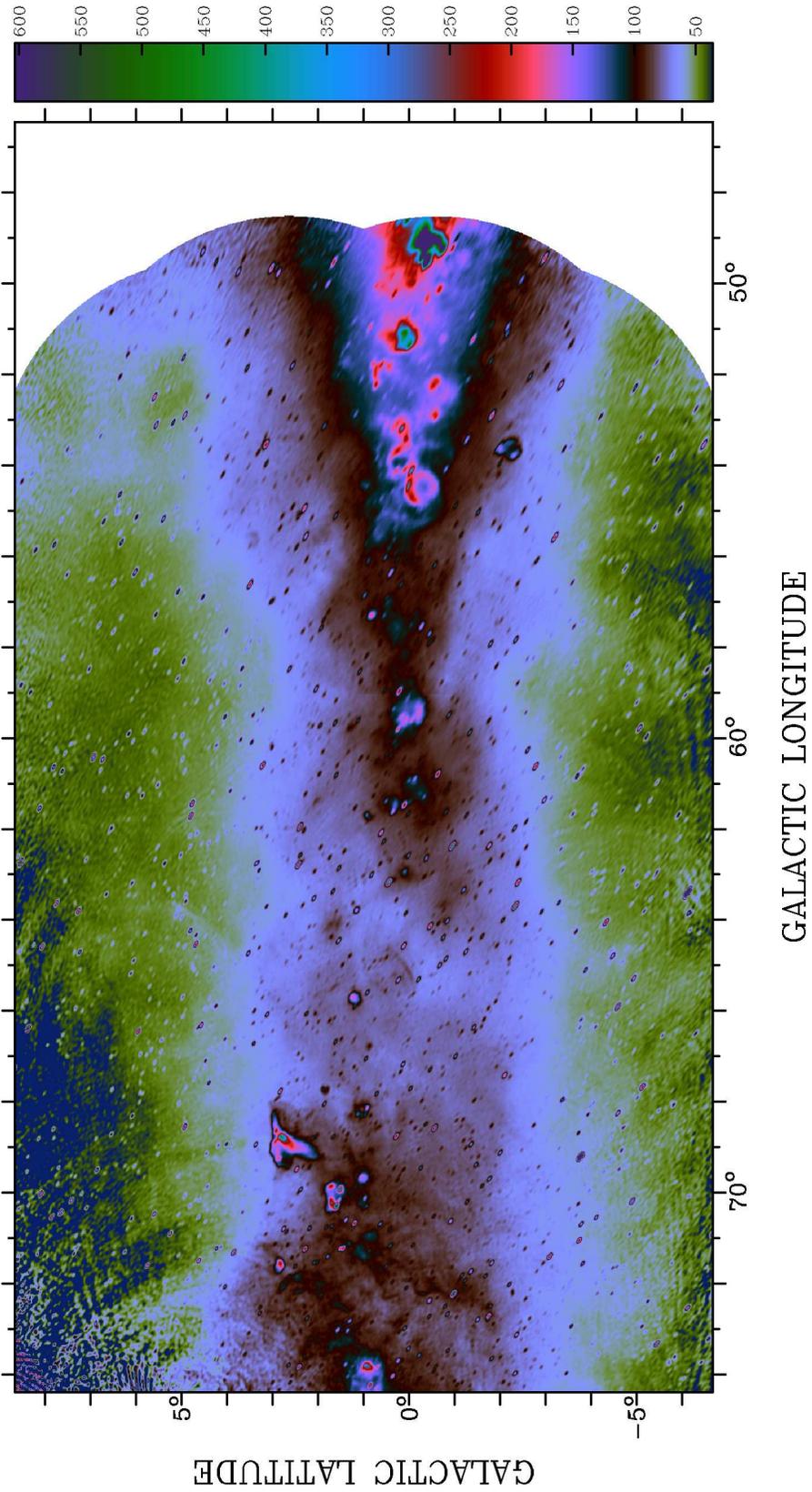}
\caption{The 408\,MHz survey image covering
${50^{\circ}\leq{\ell}\leq{74^{\circ}}}$, showing brightness temperature in
Kelvins.}
\label{survey1}
\end{center}
\end{figure*}

\begin{figure*}
\begin{center}
\includegraphics[bb = 90 35 490 765,scale=0.85,angle=0,clip=]
{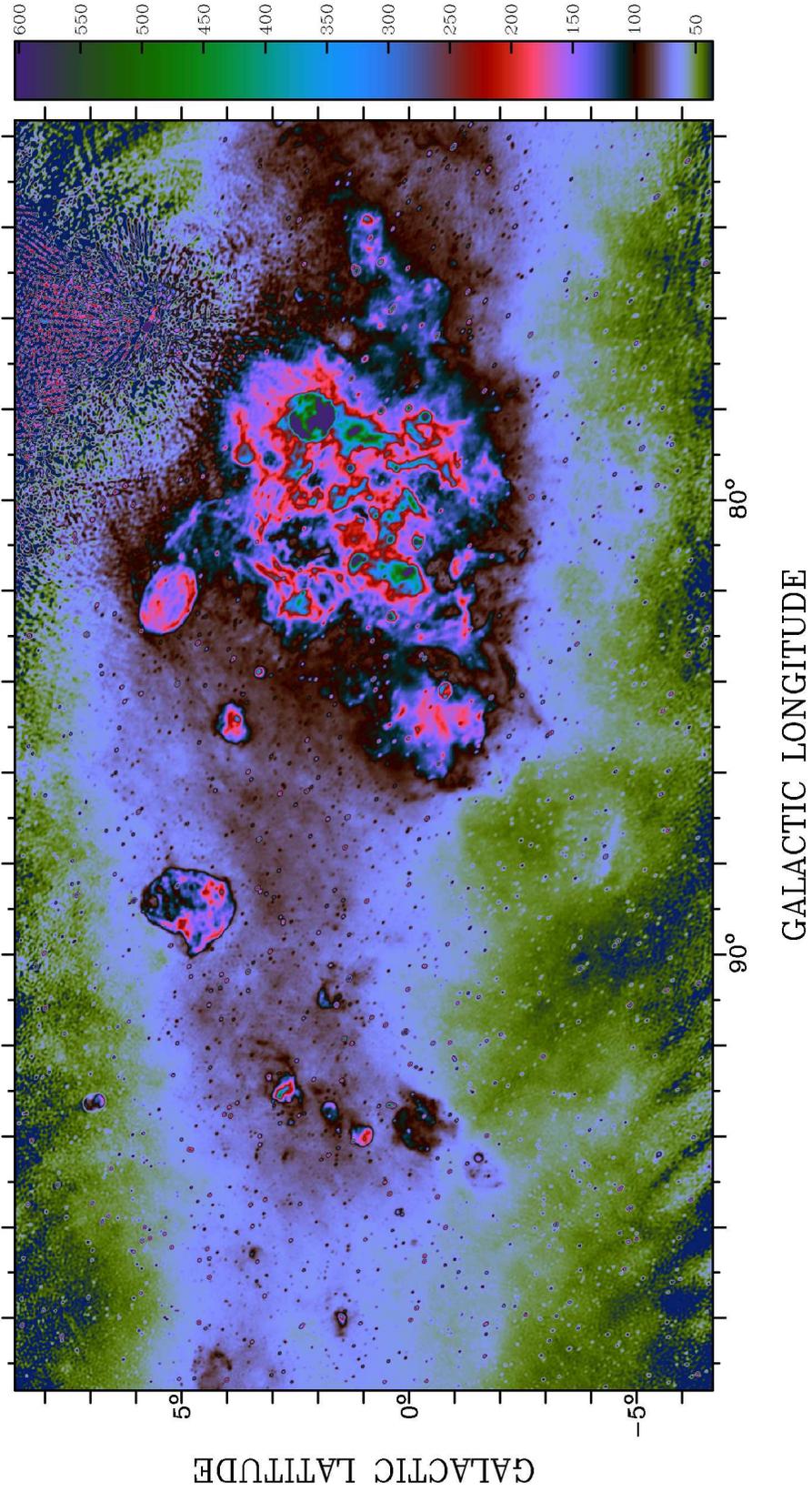}
\caption{The 408\,MHz survey image covering 
${72^{\circ}\leq{\ell}\leq{99^{\circ}}}$, showing brightness temperature in
Kelvins. Artefacts from Cygnus A are evident around the source position,
$(\ell,b)=(76.2^{\circ},5.8^{\circ})$.}
\label{survey2}
\end{center}
\end{figure*}

\begin{figure*}
\begin{center}
\includegraphics[bb = 90 35 490 765,scale=0.85,angle=0,clip=]
{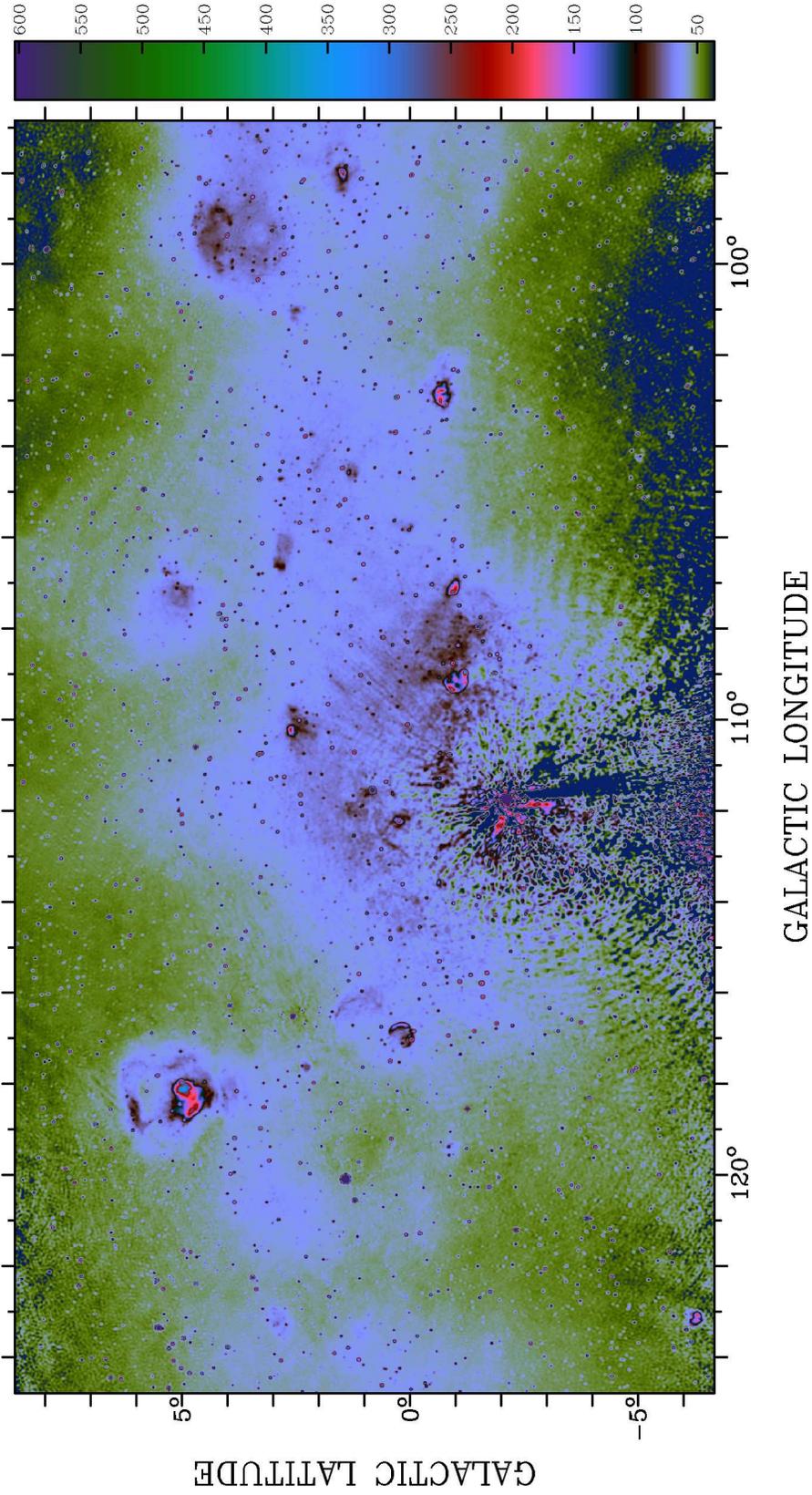}
\caption{The 408\,MHz survey image covering 
${97^{\circ}\leq{\ell}\leq{125^{\circ}}}$, showing brightness temperature in
Kelvins. Artefacts from Cassiopeia A are evident around the source position,
$(\ell,b)=(111.7^{\circ},-2.1^{\circ})$.}
\label{survey3}
\end{center}
\end{figure*}

\begin{figure*}
\begin{center}
\includegraphics[bb = 90 35 490 765,scale=0.85,angle=0,clip=]
{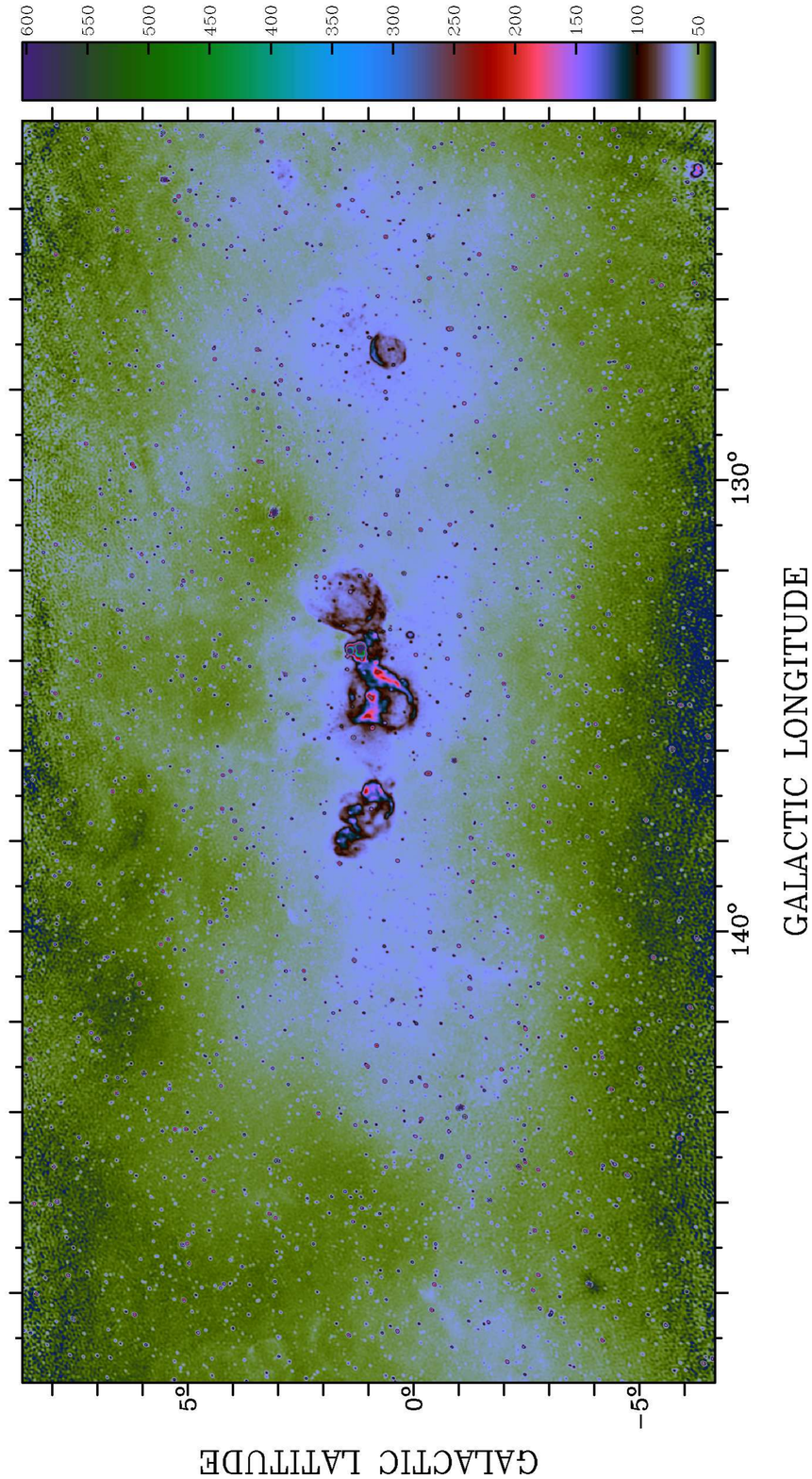}
\caption{The 408\,MHz survey image covering 
${123^{\circ}\leq{\ell}\leq{150^{\circ}}}$, showing brightness temperature in
Kelvins.}
\label{survey4}
\end{center}
\end{figure*}

\begin{figure*}
\begin{center}
\includegraphics[bb = 90 35 490 765,scale=0.85,angle=0,clip=]
{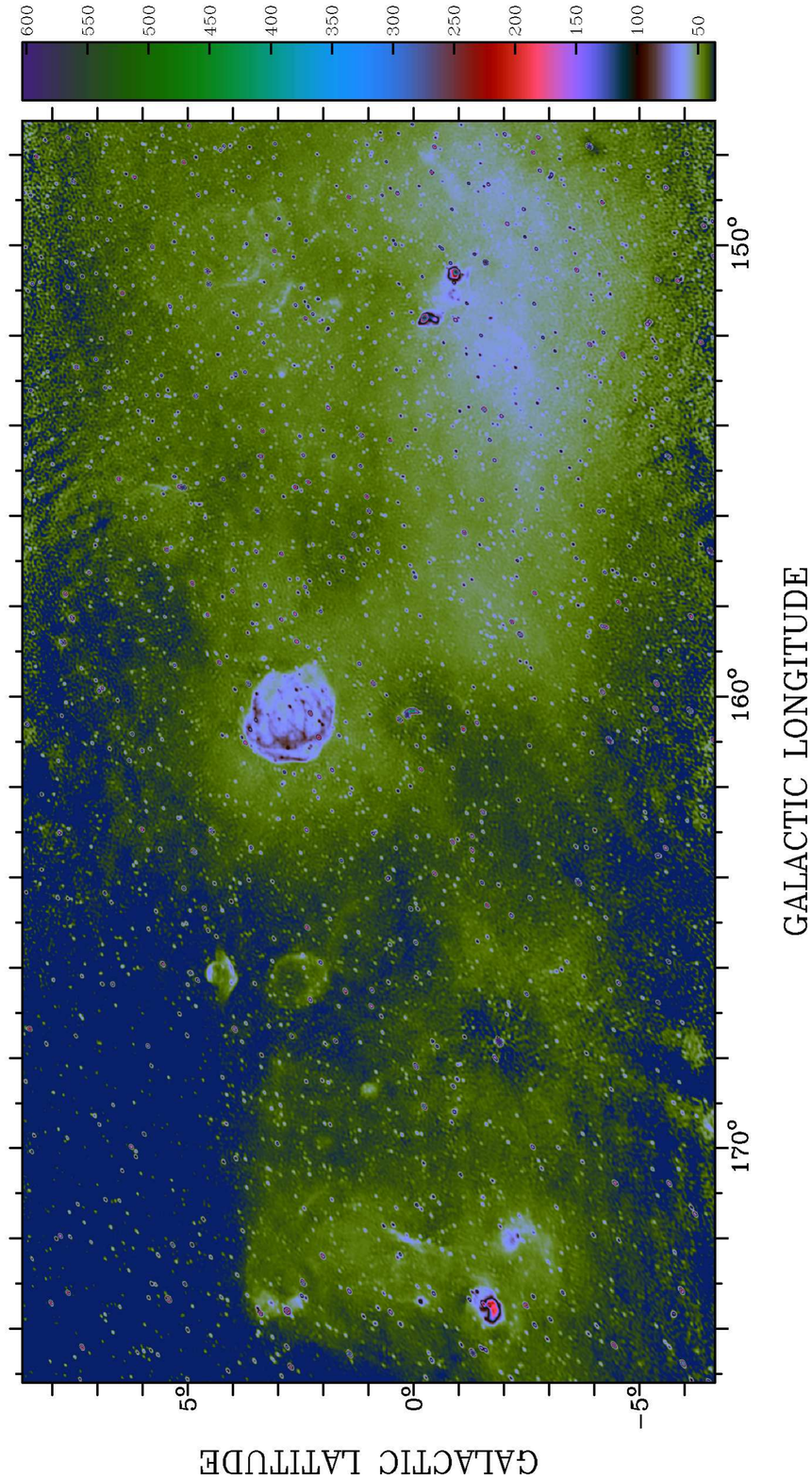}
\caption{The 408\,MHz survey image covering 
${148^{\circ}\leq{\ell}\leq{175^{\circ}}}$, showing brightness temperature in
Kelvins.}
\label{survey5}
\end{center}
\end{figure*}

\begin{figure*}
\begin{center}
\includegraphics[bb = 90 35 490 765,scale=0.85,angle=0,clip=]
{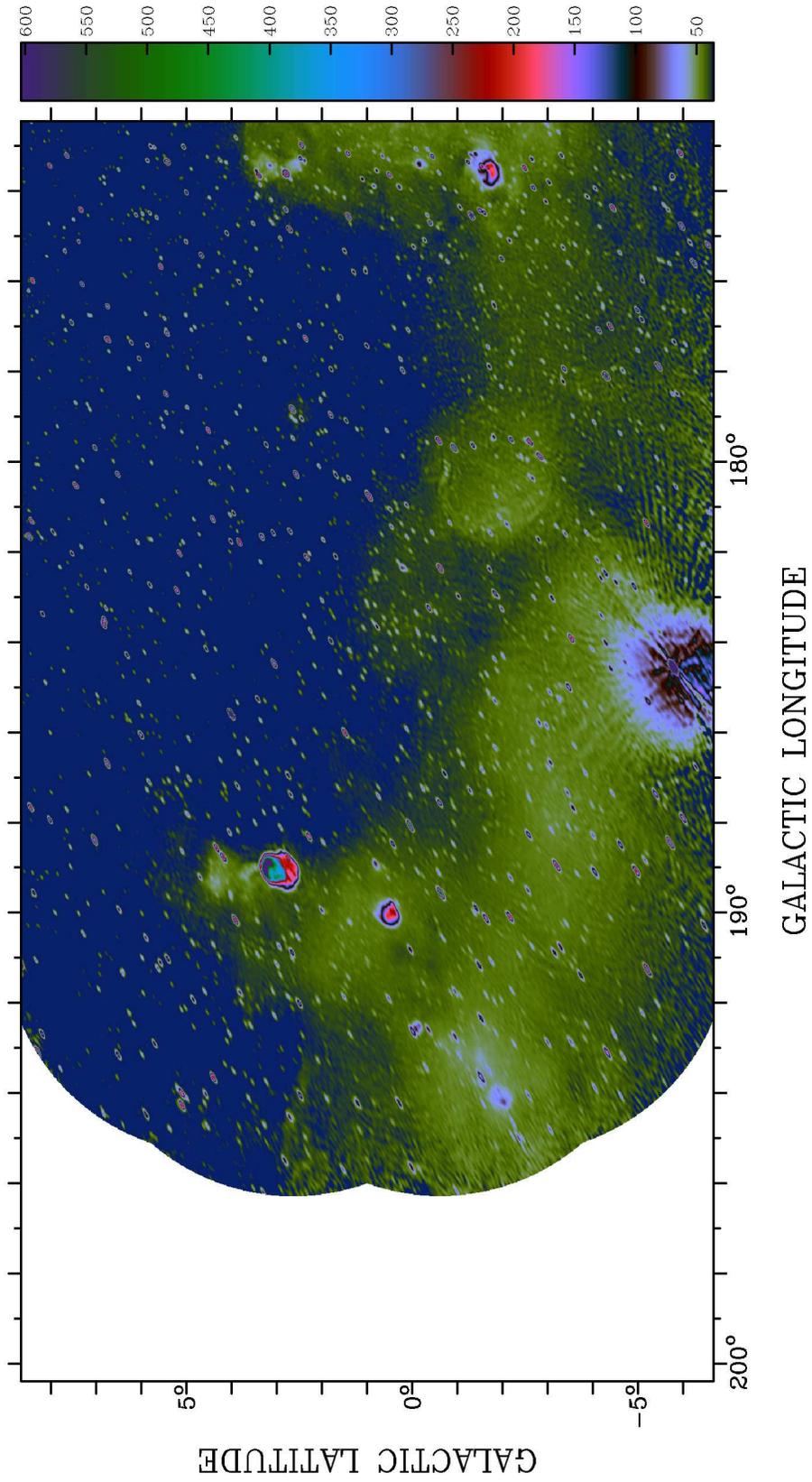}
\caption{The 408\,MHz survey image covering 
${173^{\circ}\leq{\ell}\leq{196^{\circ}}}$, showing brightness temperature in
Kelvins. Artefacts from Taurus A are evident around the source position,
$(\ell,b)=(184.6^{\circ},-5.8^{\circ})$.}
\label{survey6}
\end{center}
\end{figure*}

\begin{figure*}
\begin{center}
\includegraphics[bb = 90 35 490 765,scale=0.85,angle=0,clip=]
{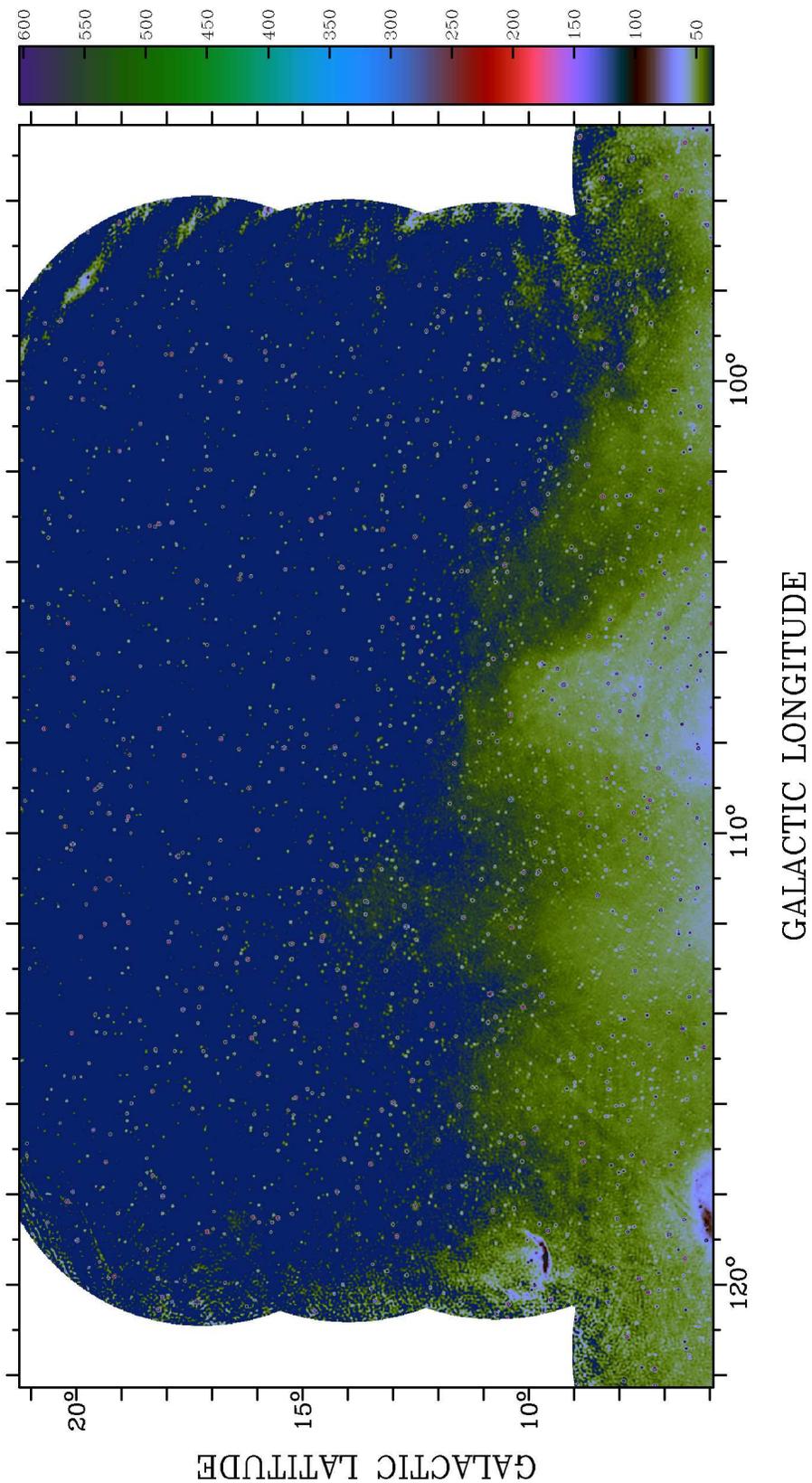}
\caption{The 408\,MHz survey image covering the high-latitude section 
${97^{\circ}\leq{\ell}\leq{120^{\circ}}}$, showing brightness temperature in
Kelvins.}
\label{survey7}
\end{center}
\end{figure*}

\section{SURVEY PRESENTATION}
\label{pres}

Figures~\ref{survey1} to~\ref{survey7} show six survey images along the Galactic
plane and one image portraying the high-latitude extension{\footnote{The survey
data described in this paper will be available at the Canadian Astronomy Data
Centre (www.cadc-ccda.hia-iha.nrc-cnrc.gc.ca/en/cgps) when the paper is
published in the Astronomical Journal}}. The color scale has been chosen with
considerable care to show the large dynamic range of the survey, but even so the
lowest level features are not always completely represented.   The data unit is
1\,K, which is equivalent to 5114 Jy/sr at 408\,MHz.

The quality of the survey images is high because of the very complete sampling
of the $(u,v)$ plane, with well-developed processing software tuned to the
properties of the telescope (\citealp{higg97,will99}), all supplemented by
strict attention to detail in image processing. Special care was taken with the
areas around the strong sources Cas~A, Cyg~A, and Tau~A. All three are within
the survey area, and image quality in their immediate surroundings is
compromised.  However, the effects are confined to relatively small areas. The
badly corrupted zones around Cas~A and Cyg~A are only $\sim5^\circ$ in extent
and that around Tau~A $\sim3^\circ$; low level effects extend to about twice
those diameters (Figures~\ref{survey1} to \ref{survey6}). Nevertheless, the
depiction of extended emission is largely unaffected.

Some of the images show striping at low levels, with the stripes spaced about
$1^\circ$ apart. These are scanning artefacts in the Haslam data that radiate
from the North Celestial Pole.  We could have used one of the ``improved''
versions of the Haslam dataset (e.g. \citealp{rema15}) where the
stripes have been reduced in amplitude by filtering in the spatial frequency
domain, but we preferred to use the original data where the sampling of the
$(u,v)$ plane was well known. There are a few small residual features from solar
interference in the DRAO ST observations, for example at
${(\ell,b)}\approx{({150^{\circ}},8^{\circ})}$.

At the two ends of the survey the telescope beam becomes highly elliptical (the
cosec\thinspace$\delta$ dependence) as is seen from the compact sources. This
does not affect representation of the extended emission.

\section{DISCUSSION}
\label{disc}

\begin{table*}
\caption{Selected surveys covering the northern Galactic plane}
\label{surveys}
\centerline{
\begin{tabular}{lcccccc}
\hline\noalign{\smallskip}
Survey     & Ref. & $F$   & Beam & \multicolumn{2}{c}{Coverage} & Sens. \\
           &      & (MHz) &      & $\ell$    & $b$ & (mK) \\ 
\noalign{\smallskip}\hline\noalign{\smallskip}
DRAO-22 & (1) & 22 & $1.1^{\circ}{\times}1.7^{\circ}$\thinspace$^a$ & 
$0^{\circ}$ to $240^{\circ}$ & \quad\quad $^b$ & \\
CGPS 408 & (2) & 408 & $2.8'$\thinspace$^a$ & 52$^{\circ}$ to 193$^{\circ}$ &
$-$6.5$^{\circ}$ to 8.5$^{\circ}$ & 760 \\
CGPS 1420 & (3) & 1420 & $58''$\thinspace$^a$ &52.5$^{\circ}$ to 192$^{\circ}$ 
& $-$3.5$^{\circ}$ to 5.5$^{\circ}$ & 60 \\
Eff-21A & (4) & 1408 & $9.4'$ & 357$^{\circ}$ to 95.5$^{\circ}$ & 
$-$4$^{\circ}$ to 4$^{\circ}$ & 40 \\
Eff-21B & (5) & 1408 & $9.4'$ & 95.5$^{\circ}$ to 240$^{\circ}$ & 
$-$4$^{\circ}$ to 5$^{\circ}$ & 30 \\
Eff-11 & (6), (7) & 2695 & $4.3'$ & 358$^{\circ}$ to 240$^{\circ}$ & 
$-$5$^{\circ}$ to 5$^{\circ}$  & 50 \\
Urumqi-A & (8) & 4800 & $9.5'$ & 10$^{\circ}$ to 60$^{\circ}$ & 
$-$5$^{\circ}$ to 5$^{\circ}$ & 1 \\
Urumqi-B & (9) & 4800 & $9.5'$ & 60$^{\circ}$ to 129$^{\circ}$ & 
$-$5$^{\circ}$ to 5$^{\circ}$ & 0.9 \\
Urumqi-C & (10) & 4800 & $9.5'$ & 129$^{\circ}$ to 230$^{\circ}$ & 
$-$5$^{\circ}$ to 5$^{\circ}$ & 0.6 \\
\noalign{\smallskip}\hline
\end{tabular}}

\thinspace References: (1) \citet{roge99}, (2) this paper, (3) \citet{tayl03},
(4) \citet{reic90a}, (5) \citet{reic97}, (6) \citet{furs90}, (7) \citet{reic90}
(8) \citet{sun11}, (9) \citet{gao10}, (10) \citet{xiao11}

\thinspace Notes: $^a$ Resolution is declination dependent - see reference for
details. $^b$ The 22\,MHz survey covers ${-28^{\circ}}<{\delta}<{80^{\circ}}$
for all Right Ascensions. 
\end{table*}

The declination limit of the telescope (${\delta}\geq{17^{\circ}}$) allows
coverage of the second quadrant of Galactic longitude with a small incursion
into the third quadrant, but with relatively little access to the inner Galaxy.
In this section we give an outline description of the regions covered by the
survey and show a few examples of the many possible uses of the data.

The images of Figures~\ref{survey1} to \ref{survey7} accurately represent all
structure from the largest angular scales to the resolution limit of the
telescope. They provide the first high-resolution view of the Galactic radio
emission in this low-frequency range. Indeed, there is no other survey of this
kind below 1.42\,GHz, the other continuum frequency of the CGPS.  In
Table~\ref{surveys} we compile a list of surveys that cover the Northern
Galactic plane. Only surveys that portray large-scale structure with good
accuracy were selected for this list, and then only the surveys with the highest
angular resolution were kept. This table emphasizes the particular value of the
two CGPS surveys, with excellent angular resolution and faithful representation
of large structure at low frequencies.

The CGPS has created unprecedented opportunities to study extended features of
the Galactic emission with sizes from a few arcminutes upwards. All other
surveys with angular resolution in the arcminute range are surveys of point
sources, with varying, but generally deficient, sensitivity to extended
structure (e.g. VLSS - \citealp{cohe07}; 7C(G) - \citealp{vess98}; WENSS -
\citealp{reng97}; Texas - \citealp{doug96}; and NVSS - \citealp{cond98}). These
are excellent surveys in their own right, but they serve another purpose. 

We note that the high resolution of our survey is essential if we are to obtain
an accurate representation of the diffuse emission. In low-resolution
surveys, discrete compact sources, and even mildly extended sources, blend
together and are impossible to separate from the diffuse emission.

\subsection{The Largest Structures}
\label{large}

Because of the low frequency of the survey we expect the large-scale diffuse
emission in the Galactic plane to be substantially non-thermal in origin. We
verify this statement, and quantify it, in Sections~\ref{cyg} and \ref{perseus}.

Sagittarius Arm emission dominates at ${\ell}\leq{60^{\circ}}$ and the
separation between the Sagittarius arm and the Local arm at
${\ell}\approx{60^{\circ}}$ is very clear. There is a concentration of discrete
sources between longitudes $59^{\circ}$ and $64^{\circ}$. \ion{H}{2} regions
Sh\,2-86, 87, 88, 89, and 97 are at distances of 2 to 3 kpc \citep{fich84}, and
cannot be part of the Sagittarius Arm, and Sh\,2-92 and 93 are at $\sim$4\,kpc,
and may be within that arm.

At longitudes up to ${\sim}70^{\circ}$ the emission peaks around
${b}={0^{\circ}}$. The Cygnus-X emission from
$73^{\circ}{\leq}{\ell}{\leq}86^{\circ}$ is centered above the mid-plane and
from ${\ell}{\approx}{86^{\circ}}$ the peak of the extended emission stays well
above ${b}={0^{\circ}}$; at  ${\ell}\approx{100^{\circ}}$ the center of the
extended emission is at ${b}\approx{3^{\circ}}$. This is the warp of the
Galactic disk \citep{binn98}. Local Arm emission dominates in the range
${68^{\circ}}\leq{\ell}\leq{100^{\circ}}$, including the large, predominantly
thermal, complex Cygnus X (${73^{\circ}}\leq{\ell}\leq{86^{\circ}},
{-3^{\circ}}\leq{b}\leq{5^{\circ}}$), discussed in Section~\ref{cyg}. Perseus Arm
emission dominates the range ${100^{\circ}}\leq{\ell}\leq{160^{\circ}}$, at
distances of 2 to 3\,kpc, with major \ion{H}{2} region complexes at intervals.
Part of the Perseus Arm is discussed in Section~\ref{perseus}.

In the anti-center, beyond ${\ell}\approx{160^{\circ}}$, we expect a blend of
Local and Perseus Arm emission. However, the diffuse emission peaks well below
the mid-plane, at ${b}\approx{-3^{\circ}}$, which may indicate that it is fairly
local in origin. Some discrete objects, on the other hand, are at higher
latitudes. The SNRs HB9 (G160.9+2.6) and VRO\,42.05.01 (G166.0+4.3) are both
Perseus Arm objects. Numerous \ion{H}{2} regions (Sh\,2-217, 219, 223, 225, 228,
231, 232, and 235 are found at Perseus arm distances \citep{fost15} in
$159^{\circ}{\leq}{\ell}{\leq}174^{\circ}$ at latitudes above ${b}={0^{\circ}}$.

We note that our survey enables us to trace the warp of the {\it{synchrotron}}
disk. The warp, at least the ISM component of it, has usually been discussed in
terms of \ion{H}{1} emission, as in \citet{binn98}. In Sections~\ref{cyg} and
\ref{perseus} we give numerical estimates of the diffuse non-thermal emission at
various places along the disk, but a complete discussion is beyond the scope of
this paper.

\subsection{Cygnus\,X and W\,80}
\label{cyg}

Massive stars shape the Galactic ISM; they enrich the ISM around them and they
trigger the formation of new stars. In this context Cygnus\,X, seen prominently
in Figure~\ref{survey2}, is an area of special interest. Cygnus\,X is a complex
region of very intense emission, most of it thermal in nature, which was long
considered to be the local spiral arm seen end-on with emission from objects
over a large range of distances superimposed ({\it{e.g.}} \citealp{wend91}).
This is partly correct, but recent evidence indicates that there are only three
major concentrations of material along the line of sight, at distances of 500 to
800\,pc, 1.0 to 1.8\,kpc, and 1.5 to 2.5\,kpc \citep{gott12}. Cyg\,OB2,
originally classified as an OB association, is now known to be much more
significant, a large cluster that contains $\sim$120 O stars
\citep{knod00,knod04}; its distance is 1.7\,kpc \footnote{\citet{rygl12}
measured parallax distances to Cygnus-X objects, and place the entire complex at
$1.40{\pm0.08}$\,kpc; their distance to W75N is $1.30{\pm0.07}$\,kpc. However,
\citet{gott12} show that the molecular gas associated with W75N is definitely
distinct from other molecular complexes.}. Cygnus\,X is one of a small number of
sites within the Galaxy where massive stars are known to have formed in great
concentrations. As the closest such site, it is an important laboratory for
study of all the pheonomena that accompany the births, lives, and deaths of the
largest stars. 

\begin{figure*}
\begin{center}
\includegraphics[bb = 105 70 550 690,angle=-90,scale=0.75,clip=]{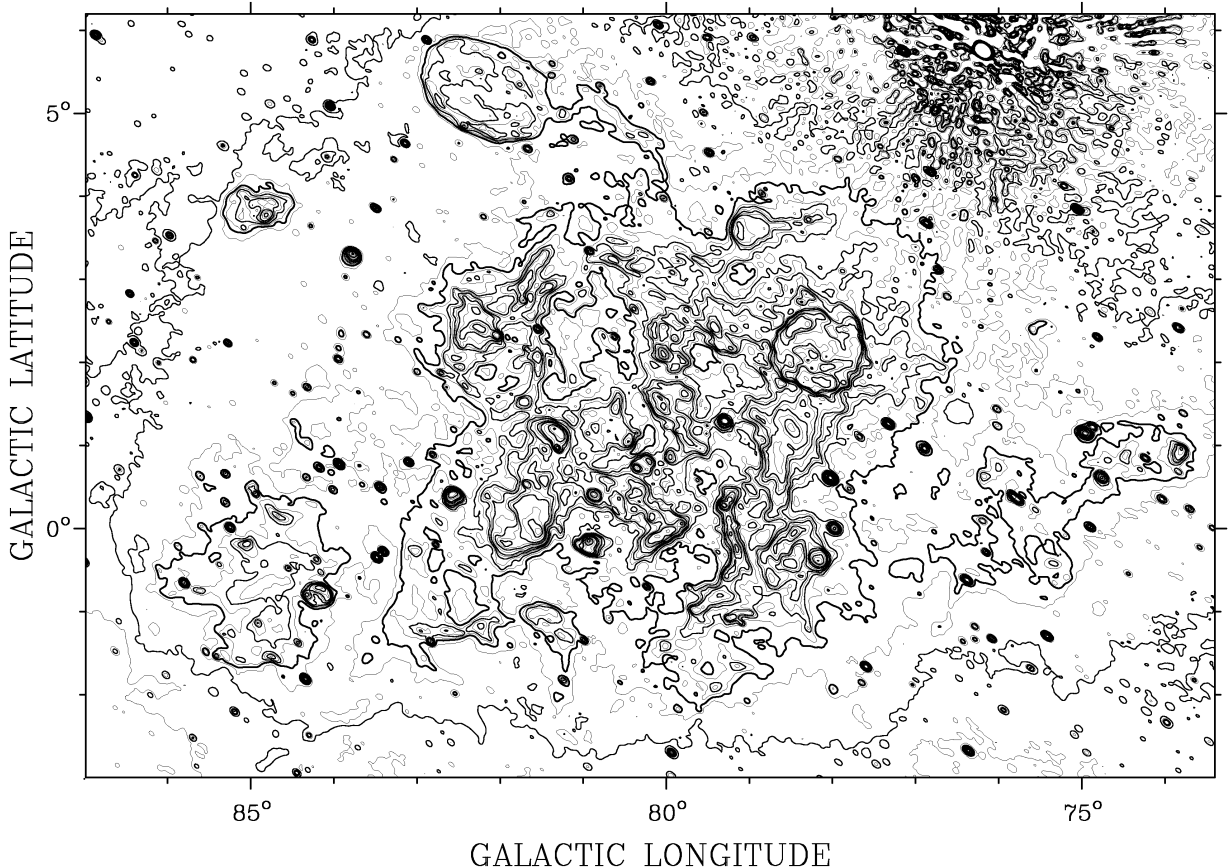}
\caption{Cygnus X and W80. Alternating light and heavy contours are used.
Contours are 60\,K to 240\,K in steps of 20\,K, 240\,K to 400\,K in steps of
40\,K, 400\,K to 800\,K in steps of 100\,K, 1000\,K and 1500\,K. The 120\,K
contour is especially heavy.}
\label{cygx}
\end{center}
\end{figure*}

W\,80, also seen in Figure~\ref{survey2}, is a prominent \ion{H}{2} region,
first recognized by \citet{west58}, that embraces two well-known optical
objects, the North American Nebula (NGC\,7000) and the Pelican Nebula
(IC\,5040). NGC\,7000 and IC\,5040 lie on either side of a conspicuous dark
cloud, L935 \citep{lynd62}, but radio observations reveal that the two optical
nebulae are simply parts of one large emission region. The whole complex is at a
distance of $550{\pm}50$\,pc \citep{laug06}, and the dominant source of
ionization is an O5V star \citep{come05}. Molecular and atomic gas in W\,80 were
mapped by \citet{feld93a} and discussed by \citet{feld93b}. \ion{H}{1}
observations show a deficiency of atomic gas over the area of W\,80, suggesting
that the bulk of the gas is in molecular form \citep{feld93b}. Molecular gas and
\ion{H}{1} self-absorption coincide, indicating the first stages of
fragmentation of the molecular clouds and the formation of young stars. Very
active star formation is found in L935 \citep{armo11}, with at least 35 HH
objects and 41 H$\alpha$ emission-line stars identified in the ``Gulf of
Mexico'' sub-region of the dark cloud. 

Figure~\ref{cygx} presents the survey data covering the Cygnus-X region and
W\,80 as a contour plot. We have used the same contour levels that were used in
the previous best data on this area, the map published by \citet{wend91}. The
resemblance to the earlier map is very strong; this is not surprising since the
earlier data were also observed with the DRAO ST. The present data have better
sensitivity by a factor of 4, superior removal of the effects of Cygnus A, and
cover a larger area.

With the improved sensitivity of the present data we see clearly the low-level
envelope that surrounds the whole of Cygnus\,X (very evident in
Figures~\ref{survey1} and \ref{survey2}, extending over the range 
${65^{\circ}}\leq{\ell}\leq{95^{\circ}}$). The emission here has thermal and
non-thermal components. First, in the data of \citet{reic88}, spectral index
between 408 and 1420\,MHz at angular resolution ${\sim}0.9^{\circ}$, this region
displays a lower temperature spectral index,
${\beta}\thinspace{\approx}\thinspace{2.4}$, than its surroundings, where
${\beta}\thinspace{\geq}\thinspace{2.6}$. Second, an extensive area of
absorption is evident in the 22\,MHz data of \citet{roge99}, extending over at
least ${68^{\circ}}\leq{\ell}\leq{90^{\circ}}$. It is likely that the thermal
component of this emission lies at a distance of about 500\,pc, the distance to
the nearer parts of Cyg\,X and of W\,80. The level of thermal emission here is
much higher than at other points along the Galactic plane. The level of
non-thermal emission behind Cygnus\,X at 408\,MHz is about 60\,K, judged by
comparing images from the present work with data from the Effelsberg 11-cm
survey (\citealp{furs90,reic90}) using the technique illustrated in
Section~\ref{perseus} for the Perseus Arm (see Figure~\ref{w345}).

\subsection{The Perseus Arm}
\label{perseus}

\begin{figure*}
\begin{center}
\includegraphics[bb = 142 80 495 650,scale=1.0,clip=]
{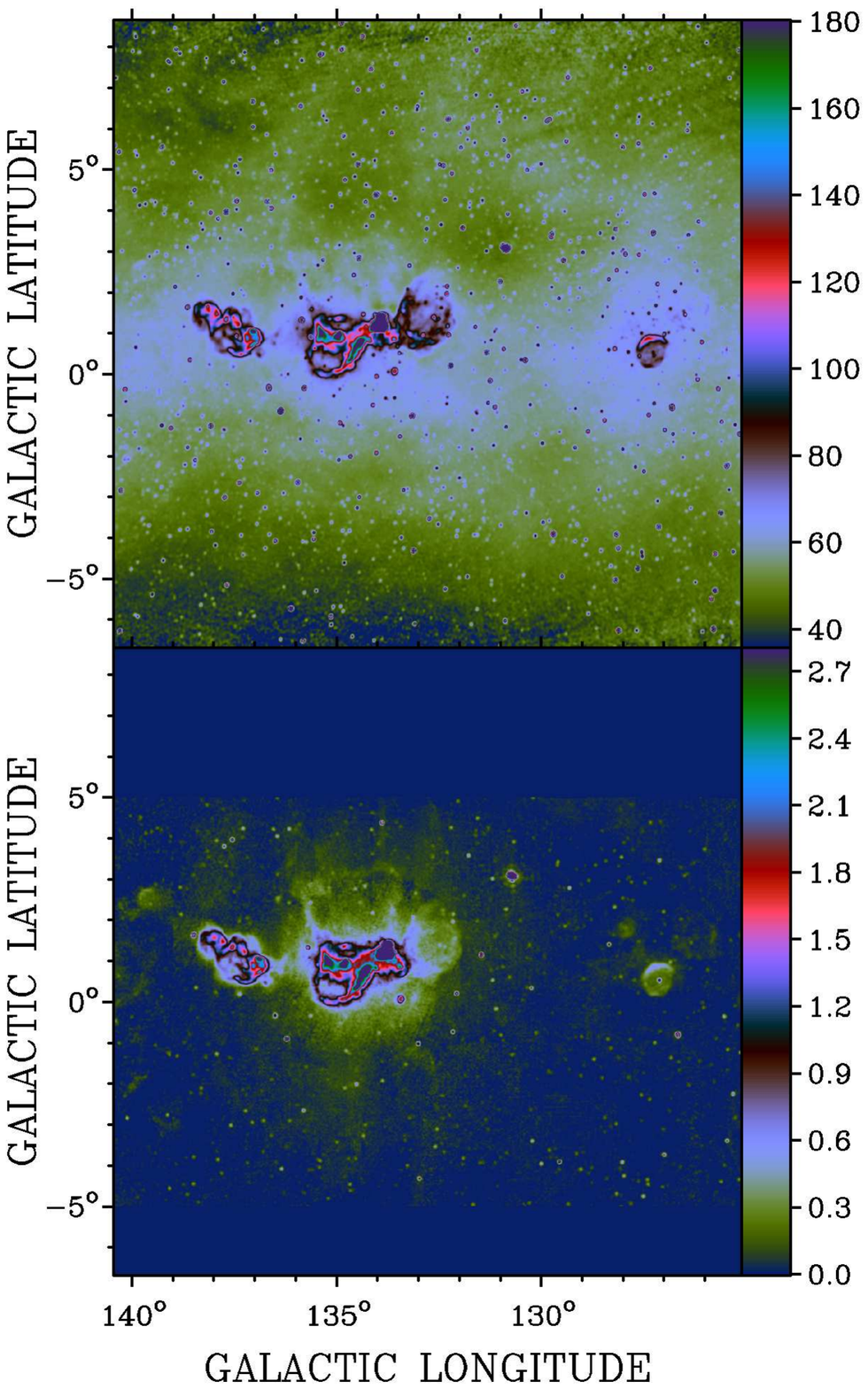}
\caption{The area around the W3/W4/W5 complex at 408\,MHz (top) and 2695\,MHz
(bottom). The angular resolution of both images is $4.3'$. The scales have been
chosen so that optically thin thermal emission has the same appearance in both
images. Note that the 2695\,MHz image does not extend beyond
${-5^{\circ}}\leq{b}\leq{5^{\circ}}$.}
\label{w345}
\end{center}
\end{figure*}

The area around the \ion{H}{2} regions W3, W4, and W5 is shown in
Figure~\ref{w345}, where data from the present survey are compared to 2695\,MHz
data from \citet{furs90} and \citet{reic90}. The 408\,MHz image has been
smoothed slightly to $4.3'$ to match the angular resolution of the 2695\,MHz
data. The color scale for the figure has been chosen so that optically thin
thermal emission has very similar appearance in both images. Non-thermal objects
will therefore stand out in the 408\,MHz image; particularly evident are HB\,3
(G132.7+0.3) and G127.1+0.5, both well known SNRs. 

The widespread low-level emission evident in the 408\,MHz image at levels up to
about 40\,K has no counterpart at 2695\,MHz. If this was thermal emission we
would expect to see it at 2695\,MHz at a level of $\sim$0.8\,K, well above the
sensitivity limit of the 2695\,MHz data. This is clearly non-thermal emission.
Considering that it is well confined to low latitudes, most of it probably
arises in the Perseus Arm. Similar extended emission, no doubt also non-thermal,
is seen at  ${\ell}\approx{150^{\circ}}$, where it reaches levels of 50 to 60\,K
(see Figure~\ref{g155}). 

\subsection{Structures of Intermediate Size}

\begin{figure*}
\begin{center}
\includegraphics[bb = 40 110 560 590,scale=0.65,clip=]
{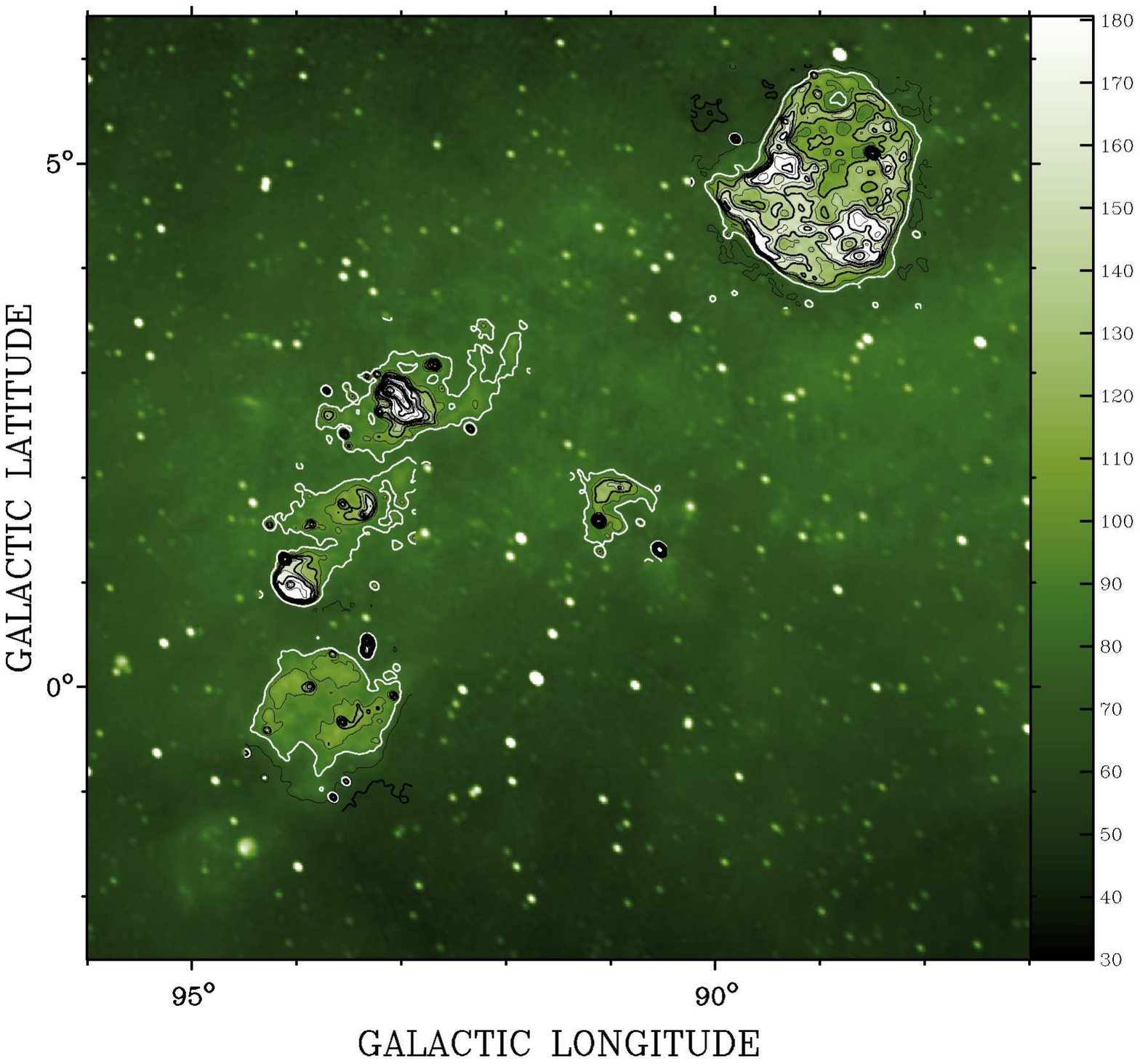}
\caption{An area $9.0^{\circ}{\times}9.0^{\circ}$ centered at 
${(\ell,b)}={(91.5^{\circ},1.9^{\circ})}$. The color scale is chosen to 
emphasize low-level extended emission. Contours are superimposed on parts of the
image around some prominent objects mentioned in the text. Contour levels are 40 to 175\,K
in steps of 15\,K, 175 to 295\,K in steps of 30\,K, 295 to 505\,K in steps of
70\,K, 600, 720, and 840\,K. the 85\,K contour is white. Compact sources of
sufficient intensity are white in the image, except where contours have been
overlaid on them and they appear black.}
\label{g91}
\end{center}
\end{figure*}

\begin{figure*}
\begin{center}
\includegraphics[bb = 35 105 555 525,scale=0.65,clip=]
{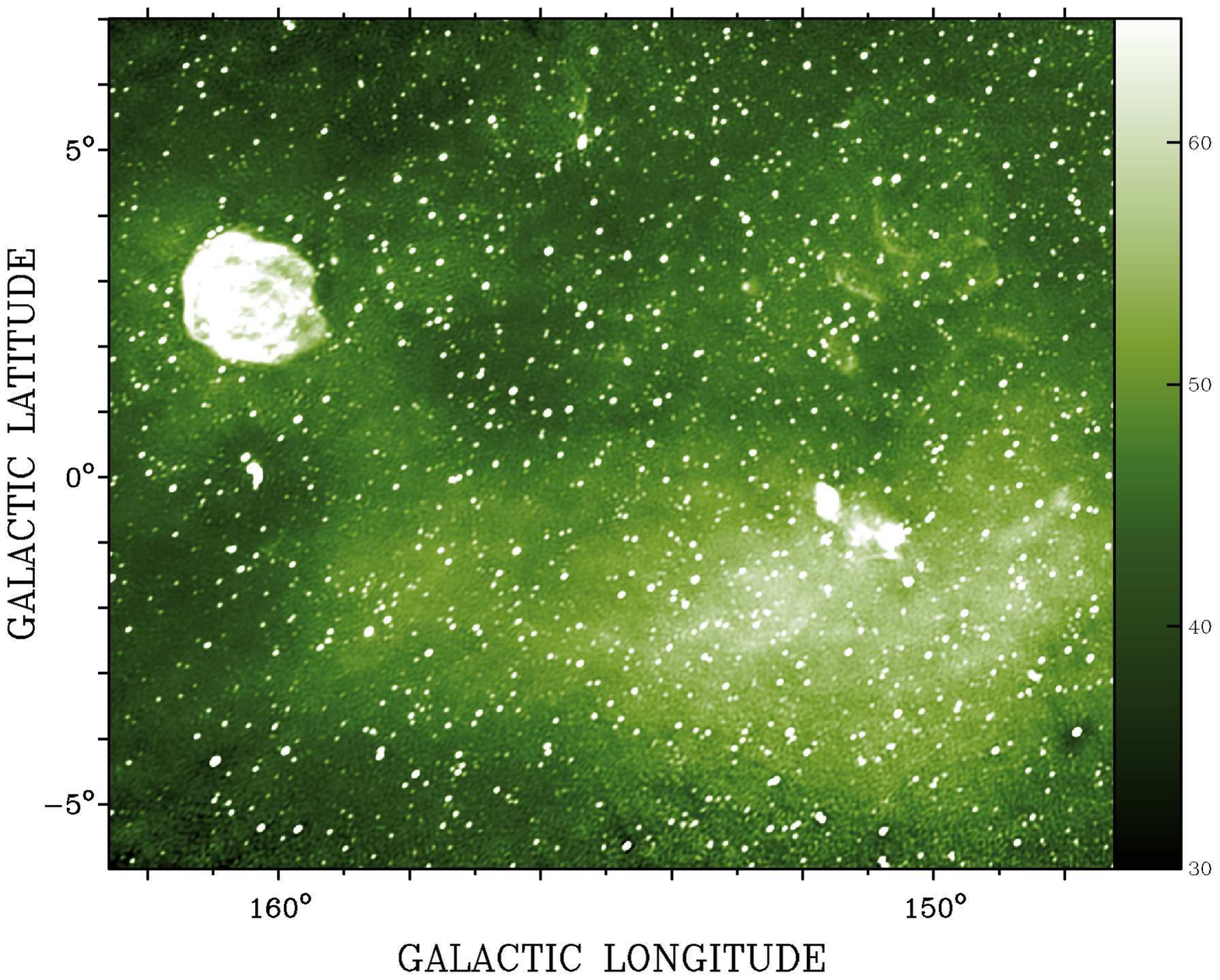}
\caption{An area $15.4^{\circ}{\times}13.0^{\circ}$ centered at 
${(\ell,b)}={(154.9^{\circ},0.9^{\circ})}$. The color scale is chosen to 
emphasize low-level extended emission. Newly discovered SNRs G149.5+3.2,
G150.5+3.8, G152.4$-$2.1, and G160.1$-$1.1 can be seen in this image. The very
faint SNRs G156.2+5.7 and G151.2+2.9, as well as the bright SNR HB9 (G160.9+2.6)
are also detectable here. Details of these objects are given in the text.}
\label{g155}
\end{center}
\end{figure*}

Figure~\ref{g91} presents a detailed look at a rich area containing a diversity
of Galactic objects over a large range of distances.  {\it{(i)}} HB21
(G89.0+4.7) is a SNR whose non-circular appearance indicates strong interaction
with the ISM. \citet{tate90} show it is colliding with a molecular cloud on its
eastern boundary; the distance to this molecular material, 0.8\,kpc, establishes
the distance to the SNR. {\it{(ii)}} CTB\,104A (G93.7$-$0.2) is a similarly
non-circular SNR, at a distance of 1.5\,kpc, that displays an unusual
phenomenon: an \ion{H}{1} shell surrounds CTB\,104A and synchrotron emitting
material has broken through the shell from the hot SNR interior, giving the SNR
its distinctly non-circular appearance \citep{uyan02}. {\it{(iii)}} CTB\,102
(G92.9+2.7), at a distance of 4.3\,kpc, is an \ion{H}{2} region and stellar wind
bubble of size 100 to 130\,pc \citep{arvi09}. {\it{(iv)}} The 408~MHz emission
from BG2107+49 (G91.0+1.7) is thermal, but the source has a head-tail
appearance, reminiscent of a radio galaxy. The tail traces the outline of a
large, old stellar-wind bubble, while the head is relatively young \ion{H}{2}
region \citep{vand90}, probably recently formed within the shell. The physical
size of the bubble is 150\,pc and the distance $\sim$10\,kpc. {\it{(v)}} NRAO655
(G93.4+1.8) and 3C\,434.1 (G94.0+1.0) are an \ion{H}{2} and SNR respectively,
but the two are at very similar distances (\citealp{fost01,fost04}) and may be
parts of a larger complex. Newer distance estimates place the entire complex at
6.2\,kpc, in the Outer Arm of the Galaxy (T. Foster, private communication,
2017). {\it{(vi)}} Sharpless 2-124 (G94.6$-$1.5) is an \ion{H}{2} region at a
distance of 3.78\,kpc \citep{fost15}. The high angular resolution of this survey
has enabled us to separate these objects in this crowded part of the sky.

At 408\,MHz, among the sources of intermediate size, supernova remnants (SNRs)
are dominant, more obvious than \ion{H}{2} regions.  The excellent sensitivity
of the survey to sources of synchrotron emission in the intermediate range of
sizes is demonstrated in Figure~\ref{g155}, showing an area of the Galactic
plane in the second quadrant. This field is particularly rich in SNRs, and
emission from seven SNRs is evident here. The brightest is HB9 (G160.9+2.6) and
the faintest is G156.2+5.7, bright in X-rays \citep{asch91} but extremely faint
at radio wavelengths \citep{reic92}. SNRs discovered from analysis of the data
described here are also evident in Figure~\ref{g155}, namely G151.2+2.9
\citep{kert07}, G149.5+3.2, G150.5+3.8, and G160.1$-$1.1 \citep{gerb14} and
G152.5$-$2.1 \citep{fost13}. Other SNR discoveries in the CGPS data were
reported by \citet{koth01}, \citet{koth03}, \cite{koth05}, \citet{tian07}, and
\citet{koth14}. The key parameter in these discoveries has been spectral index,
underlining the importance of a high-resolution survey at a low frequency. The
power of low-frequency imaging with good angular resolution to discriminate
between non-thermal and thermal emission is well illustrated by the work of
\citet{fost06}. These authors used the 408\,MHz data, with other data, to show
that OA\,184 (G166.2+2.5), long classified as a SNR, is actually an \ion{H}{2}
region. \citet{koth06} present a catalog of SNRs covering 70\% of the CGPS area.

The other surveys listed in Table~\ref{surveys} are invaluable resources in work
like this, but the two CGPS surveys are crucial: they provide two surveys with
high angular resolution with accurate depiction of extended emission. The CGPS
1420\,MHz polarization data \citep{land10} have been especially valuable for
recognition of SNR emission. Discovery and analysis of extended
objects in the survey data have been accomplished by subtracting compact sources
and smoothing the resulting map. This process is limited by source
confusion, not thermal noise, as can be seen from a close inspection of
Figure~\ref{g155}. 

\section{CONCLUSIONS}
\label{concl}

We have described the execution and the data processing for a survey of the
radio emission from the Galactic plane at 408\,MHz.  We have presented the
survey images, and have given details of electronic access to them. Complete
sampling of all spatial scales from the largest to the resolution limit 
($2.8' \times 2.8'$\,cosec\thinspace$\delta$) has been achieved
by combining data from aperture-synthesis and single-antenna telescopes. The
image quality, especially the representation of extended structure, is extremely
high.  This is the only survey of the Galactic plane below 1\,GHz that has
arcminute resolution and faithful rendering of large structure. We have
illustrated a number of uses of the survey data which exploit its good
resolution; some allow us to reach conclusions about the extended emission. In
particular we have traced the warp in the synchrotron component of the disk in
the outer Galaxy, and we have shown that the non-thermal contribution along the
Galactic plane at 408\,MHz amounts to $\sim$60\,K at ${\ell}\approx{80^{\circ}}$
and $\sim$40\,K at ${\ell}\approx{135^{\circ}}$.

We have established a flux density scale at 408\,MHz by careful selection of
calibration sources from three extensive source catalogs, the NVSS, the
VLSS, and the 365\,MHz Texas survey. This work supersedes previous attempts to
calibrate the CGPS 408-MHz survey. The accuracy of flux densities is estimated
to be 6\%.  The survey is partly limited by thermal noise, but also by source
confusion. A complete catalog of small-diameter sources in the survey area will
be presented in a future paper.

\acknowledgments

The contributions of many people underpin the work that we are privileged to
report here: the entire CGPS team, at DRAO and in Canadian universities, played
significant roles. We are grateful to Tyler Foster who provided valuable input
on Galactic structure and distances to individual objects. We wish to record
that Grote Reber contributed funds and encouragement to the realization of the
408-MHz channel on the Synthesis Telescope. The Dominion Radio Astrophysical
Observatory is a National Facility operated by the National Research Council
Canada. The Canadian Galactic Plane Survey is a Canadian project with
international partners, and is supported by the Natural Sciences and Engineering
Research Council (NSERC).

\bibliographystyle{aa} 
\bibliography{arxiv2} 

\begin{table}
\centering
\resizebox{500pt}{40pt}
{\begin{tabular}{|l|l|l|r|r|r|r|r|r|r|r|r|r|}
\hline
  \multicolumn{1}{|c|}{Name} &
  \multicolumn{1}{c|}{Right Ascension(HMS)} &
  \multicolumn{1}{c|}{Declination(DMS)} &
  \multicolumn{1}{c|}{S408(Jy)} &
  \multicolumn{1}{c|}{$\sigma$\_S408(Jy)} &
  \multicolumn{1}{c|}{S1420(Jy)} &
  \multicolumn{1}{c|}{$\sigma$\_S1420(Jy)} &
  \multicolumn{1}{c|}{S365(Jy)} &
  \multicolumn{1}{c|}{$\sigma$\_S365(Jy)} &
  \multicolumn{1}{c|}{S74(Jy)} &
  \multicolumn{1}{c|}{$\sigma$\_S74(Jy)} &
  \multicolumn{1}{c|}{S1000(Jy)} &
  \multicolumn{1}{c|}{$\sigma$\_S1000(Jy)}  \\
\hline
NVSS000041+391804 & 00 00 41.51 & +39 18 04.5 & 0.400 & 0.011 & 0.2072 & 0.0062
& 0.437 & 0.030 & 0.92 & 0.14 & 0.250 & 0.004 \\
NVSS000045$-$272251 & 00 00 45.63 & $-$27 22 51.5 & 0.795 & 0.048 & 0.2343 
& 0070 & 0.956 & 0.049 & 3.74 & 0.38 & 0.333 & 0.016 \\
NVSS000104+101928 & 00 01 04.57 & +10 19 28.3 & 0.736 & 0.011 & 0.3202 & 0.0096
& 0.807 & 0.043 & 2.23 & 0.23 & 0.406 & 0.005 \\
\hline
\hline
  \multicolumn{1}{|c|}{Name} &
  \multicolumn{1}{c|}{Right Ascension(HMS)} &
  \multicolumn{1}{c|}{Declination(DMS)} &
  \multicolumn{1}{c|}{alpha} &
  \multicolumn{1}{c|}{$\sigma$\_alpha} &
  \multicolumn{1}{c|}{cov(S1000,alpha)} &
  \multicolumn{1}{c|}{$\chi^2$} &
  \multicolumn{1}{c|}{$\sigma$} \\
\hline
NVSS000041+391804 & 00 00 41.51 & +39 18 04.5 & 
$-$0.542 & 0.024 & 0 & 0.3829 & 0.86 & & & & &\\
NVSS000045$-$272251 & 00 00 45.63 & $-$27 22 51.5 &
$-$0.969 & 0.053 & 0 & 3.449 & 2.41 & & & & &\\
NVSS000104+101928 & 00 01 04.57 & +10 19 28.3 &
$-$0.665 & 0.013 & 0 & 0.188 & 0.59 & & & & &\\
\hline
\end{tabular}}
\caption{Example Entries of the Calibration Source List}
\label{table:cal_src}
\end{table}

\end{document}